\newif\ifShowKeys
\ShowKeystrue

\documentclass[11pt,a4paper]{article} 
\usepackage[height=8.8in,width=6.45in]{geometry}

\usepackage[hidelinks]{hyperref}

\usepackage{tocloft}

\usepackage{amsmath, amssymb,amsthm}
\usepackage{amsthm}
\numberwithin{equation}{section}

\usepackage{bm,environ,mathrsfs,array,arydshln}
\usepackage{booktabs,float,slashed,hyperref}
\usepackage[mathcal]{euscript}
\usepackage{tensor} 						
\usepackage{mathabx}
\usepackage{simpler-wick}

\usepackage{aurical}
\usepackage[T1]{fontenc}

\usepackage[nodayofweek]{date time}
\usepackage{graphicx,epsfig,epic}
\usepackage{tikz} 
\usepackage{tikzfeynman}
\usetikzlibrary{arrows,decorations.pathreplacing,decorations.markings,snakes}
\tikzset{middlearrow/.style={decoration={markings, mark= at position 0.5 with {\arrow{#1}} ,
}, postaction={decorate}}}
\tikzset{decoration={snake,amplitude=.4mm,segment length=2mm,
                       post length=0mm,pre length=0mm}}

\usepackage{framed}						
\definecolor{shadecolor}{rgb}{0.95,0.95,0.97}

\usepackage{comment}

\allowdisplaybreaks

\definecolor{myred}{RGB}{233, 33, 45}

\newcommand{\bs}{\begin{shaded}}
\newcommand{\es}{\end{shaded}}
\def\ba#1\ea{\begin{align}#1\end{align}}		
\newcommand{\be}{\begin{equation}}
\newcommand{\ee}{\end{equation}}
\newcommand{\mc}{\mathcal }

\newcommand{\la}{\label}
\newcommand{\eps}{\varepsilon}

\newcommand{\lp}{\notag \\ & }

\DeclareMathOperator{\tr}{\text{tr}}

\newcommand{\wt}{\widetilde}

\newcommand{\cf}{\textit{cf.} }
\newcommand{\ie}{\textit{i.e.} }

\newcommand{\N}{\mathcal N}

\makeatletter
\DeclareFontFamily{OMX}{MnSymbolE}{}
\DeclareSymbolFont{MnLargeSymbols}{OMX}{MnSymbolE}{m}{n}
\SetSymbolFont{MnLargeSymbols}{bold}{OMX}{MnSymbolE}{b}{n}
\DeclareFontShape{OMX}{MnSymbolE}{m}{n}{
<-6>  MnSymbolE5
   <6-7>  MnSymbolE6
   <7-8>  MnSymbolE7
   <8-9>  MnSymbolE8
   <9-10> MnSymbolE9
  <10-12> MnSymbolE10
  <12->   MnSymbolE12
}{}
\DeclareFontShape{OMX}{MnSymbolE}{b}{n}{
<-6>  MnSymbolE-Bold5
   <6-7>  MnSymbolE-Bold6
   <7-8>  MnSymbolE-Bold7
   <8-9>  MnSymbolE-Bold8
   <9-10> MnSymbolE-Bold9
  <10-12> MnSymbolE-Bold10
  <12->   MnSymbolE-Bold12
}{}

\let\llangle\@undefined
\let\rrangle\@undefined
\DeclareMathDelimiter{\llangle}{\mathopen}%
 {MnLargeSymbols}{'164}{MnLargeSymbols}{'164}
\DeclareMathDelimiter{\rrangle}{\mathclose}%
 {MnLargeSymbols}{'171}{MnLargeSymbols}{'171}
\makeatother

\catcode`,\active

\catcode`\,12

\def\XXint#1#2#3{{\setbox0=\hbox{$#1{#2#3}{\int}$}
     \vcenter{\hbox{$#2#3$}}\kern-.5\wd0}}

\newcommand{\vev}[1]{\langle  #1 \rangle}

\newcommand{\sql}{\sqrt\lambda}

\newcommand{\z}{\zeta}

\def \ov {\over}
\def \ci {\cite}
\def \foot {\footnote}
\def \N {{\cal N}}
\def \b{\beta}

\def \del{\partial}
\def \p {\phi}

\newcommand{\rf}[1]{(\ref{#1})}

\def \l {\lambda}
\def \iffa {\iffalse}
 
\def \a  {\alpha}
\def \lan  {\langle}   \def \ran {\rangle}
\def \W {W^{(0)}}

\def \ve  {\varepsilon}
\def \OO {{\cal O}}
\def \tr {{\rm Tr\,}}

\def \g  {\gamma}\def \te {\textstyle} 

\def \sql {{\sqrt \l}}  \def \D   {\Delta}

\def \z   {\zeta}

\def \WZ {W^{(\z)}}
\def \lla {\llangle}
\def \rra {\rrangle}
\def \no {\nonumber}

\def \t {\tau}

   \def \N  {{\cal N}}     
 \def \C  {{\cal C}}

\def \eps {\epsilon}

\def \PP{{\cal P}}

\def \PP  {\text{P}}

\begin{document}

\begin{titlepage}


\begin{tabbing}
\hspace*{11.5cm} \=  \kill 
\>  Imperial-TP-AT-2021-04  \\
\> 
\end{tabbing}

\vspace*{15mm}
\begin{center}
{\LARGE\sc  Higher order RG flow}
\vskip 5pt
{\LARGE\sc   on the Wilson line in $\N=4$ SYM}

\vspace*{10mm}

{\Large M. Beccaria${}^{\,a}$, S. Giombi$^{\,b}$, A.A. Tseytlin$^{\,c, }$\footnote{\ Also at the Institute for Theoretical and Mathematical Physics (ITMP) of Moscow University   and Lebedev Institute.}} 

\vspace*{4mm}
	
${}^a$ Universit\`a del Salento, Dipartimento di Matematica e Fisica \textit{Ennio De Giorgi},\\ 
		and I.N.F.N. - sezione di Lecce, Via Arnesano, I-73100 Lecce, Italy
			\vskip 0.3cm
${}^b$  Department of Physics, Princeton University, Princeton, NJ 08544, USA
			\vskip 0.3cm
${}^c$ Blackett Laboratory, Imperial College London SW7 2AZ, U.K.
			\vskip 0.3cm
			
\vskip 0.2cm
	{\small
		E-mail:
		\texttt{matteo.beccaria@le.infn.it},\ \texttt{sgiombi@princeton.edu}, \  \texttt{tseytlin@imperial.ac.uk}
	}
\vspace*{0.8cm}
\end{center}

\begin{abstract}  
\noindent
Extending earlier work,  we  find  the  two-loop term in the beta-function  for the scalar coupling  $\zeta$ in a generalized   Wilson loop operator of the  ${\cal N}=4$  SYM theory, working in the  planar weak-coupling expansion. The beta-function for  $\zeta$ has fixed points at $\zeta=\pm 1$ and $\zeta=0$, corresponding respectively to the supersymmetric Wilson-Maldacena loop and to the standard Wilson loop without scalar coupling. As a consequence of our result for the beta-function,  we  obtain a prediction for the two-loop term in the anomalous dimension of the scalar field 
inserted on the standard Wilson loop.  We  also  find  a  subset of  higher-loop  contributions
 (with highest powers of  $\zeta$   at each order in `t Hooft coupling $\lambda$) 
coming from the scalar ladder graphs  determining  the corresponding terms in  the five-loop 
beta-function.  We   discuss  the  related  structure  of  the  circular Wilson loop expectation value
commenting, in particular,  on  consistency with a 1d defect version of the F-theorem. We   also 
compute  (to two loops in the planar ladder  model approximation)   the two-point  correlators  of scalars inserted  on  the  Wilson  line.

\iffa 
Extending earlier work,  we  find  the  two-loop term in the beta-function  for the scalar coupling  $\zeta$ in a generalized  
Wilson loop operator of the  $\N=4$  SYM theory, working in the  planar weak-coupling expansion. The beta-function for 
$\zeta$ has fixed points at $\z=\pm 1$ and $\z=0$, corresponding respectively to the supersymmetric Wilson-Maldacena loop and to
the standard Wilson loop without scalar coupling.
As a consequence of our result for the beta-function,  we  obtain a prediction for the two-loop term in the anomalous dimension of the scalar field 
inserted on the standard Wilson loop. 
We  also  find  a  subset of  higher-loop  contributions
 (with highest powers of  $\z$   at each order in $\l$) 
coming from the scalar ladder graphs 
 thus determining  the corresponding terms in  the five-loop   beta-function. 
We   discuss the  related  structure  of  the  circular Wilson loop expectation value
(commenting, in particular,  on  consistency with a 1d defect version of the F-theorem), 
as well as of  the two-point  correlators  of scalars inserted  on  the loop. 
\fi
\end{abstract}
\vskip 0.5cm
	{
	}
\end{titlepage}

\tableofcontents
\vspace{1cm}

\def \W  {{\rm W}}
\def \ed {
\bibliography{BT-Biblio}
\bibliographystyle{JHEP}
\end{document}
}

\def \Do  {\Delta^{(1)}}
\def \Dz  {\Delta^{(0)}}
\def \Dze  {\Delta^{(\z)}}

\def \GW {{\cal W}}

\section{Introduction  and summary}

The   expectation value of the 
Wilson loop  (WL)  operator $\lan  \tr\PP e^{i\oint A}\ran $    is a key observable in gauge theories. In supersymmetric theories, it has a 
 ``locally supersymmetric'' analog   obtained  by adding a particular  coupling to extra fields in the gauge field multiplet.
The prototypical example is the Wilson-Maldacena loop (WML)  \cite{Maldacena:1998im,Rey:1998ik} in the
$\N=4$ Super Yang-Mills (SYM) theory 
which contains  an extra scalar coupling, and as a result has a  particularly simple structure
controlled by the underlying supersymmetry.  When the Wilson loop contour is a circle or a straight line, both the WL and WML preserve a one-dimensional 
conformal symmetry $SL(2,\mathbb{R})$ and may be regarded as 1d conformal defects of the $\N=4$ SYM theory. A circular or straight WML also globally preserves half of 
the superconformal symmetry.  

 For a generic smooth contour, the standard WL in YM  theory 
 is known to be  renormalizable: power divergences 
  exponentiate and factorize   (or simply absent in dimensional regularization)
  while  the logarithmic divergences  disappear after the renormalization of gauge coupling 
\cite{Polyakov:1980ca,Dotsenko:1979wb,Gervais:1979fv,Arefeva:1980zd,Dorn:1986dt,Marinho:1987fs}. The latter are absent in $\N=4$ SYM  so  both WL and WML 
have finite expectation values which in the planar limit are given by the   functions of the 't Hooft coupling $\l= g^2_{\rm YM} N$. 

 It is of interest  to study a family  of  more general  Wilson loop operators that interpolate 
   between 
  the  WML   and the standard WL  \cite{Alday:2007he}. This  one-parameter family   was introduced in \cite{Polchinski:2011im}   and further studied in \ci{Beccaria:2017rbe,Beccaria:2018ocq}. It is of interest, in particular, in the context of the 1d defect QFT  interpretation (see, in particular,    \ci{Cooke:2017qgm,Giombi:2017cqn, Liendo:2018ukf,Beccaria:2019dws,Correa:2019rdk,Agmon:2020pde,Cuomo:2021rkm}).  Explicitly, this generalized Wilson loop 
operator depends on an arbitrary coefficient $\z$ in front of the 
 coupling to the 6 scalars $\p_m$ and so it interpolates between the standard  the WL  ($\z=0$) and the WML ($\z=1$)  cases~\ci{Polchinski:2011im}
\be
\la{1.1}
W^{(\z)}  (C) = \frac{1}{N}\,\text{Tr}\,\PP\,\exp\oint_{C}d\tau\,\Big[i\,A_{\mu}(x)\,\dot x^{\mu}
+ \z \phi_{m}(x) \, \theta^{m} \,|\dot x|\,\Big], \qquad\qquad  \theta_{m}^{2}=1\ . 
\ee
We   may     choose   the  unit 
vector  $\theta_m$  to be along 6-th direction, \ie $\phi_m \theta^m =\phi_6\equiv \phi$. 
The  expectation value
 $\vev{\WZ}$  for a smooth contour  $C$ will 
  have   logarithmic divergences that can be absorbed into 
   a renormalization of  the coupling $\z$. 
   Then  
  the renormalized  value  of  $\vev{\WZ}$   will   be given by (in the planar limit)  
 \be \la{1.2}
\vev{\WZ}  \equiv \W\big(\l; \z(\mu), \mu\big)   
 \ , \ \ \ \  \qquad   \mu {\del \ov \del \mu}  \W  + \beta_\z  { \del \ov \del \z} \W =0  \ , \qquad \ \ 
 \beta_\z =\mu  { d \z \ov d  \mu } \ , 
\ee
 where   $\mu$ is  a renormalization scale.\footnote{For a specific contour $\W$ will  depend on 
  $\mu$ (that has dimension of a mass)   in combination with some effective length
 characterizing the loop geometry, like radius for a circular loop.}  The running of $\zeta$ then defines 
a 1d RG flow  between the WL and WML operators.  Since $\p_m \to -\p_m$ is a symmetry of the SYM path integral, 
  $\W$   should be invariant under $\z \to -\z$. In what follows  we shall assume that $\z\geq  0$. 
  In the planar weak coupling expansion the  leading order  term   in the    beta-function  was found in 
 \ci{Polchinski:2011im}  to be 
 \be  
\beta_\z = - {\l \ov 8 \pi^2}  \z (1-\z^2)   + \mc O  ( \l^2)   \ .  \la{1.3} \ee
The WL ($\z=0$) and WML  ($\z=1$) cases  in \rf{1.1} 
are  expected to be  the  only two  fixed points  also at higher orders in $\l$.\foot{While 
we  shall   mostly discuss  only planar  contributions    let us mention that  in general 
 the coefficient  $N$  in  $\l$  \rf{1.3}  for a general  simple group 
  is  the value of  quadratic Casimir in the adjoint representation
    (and thus does not depend on the representation 
  of the Wilson loop).
  For  $U(N)$ the  coefficient  in \rf{1.3}    is  to be   multiplied   by  $(1- 1/N^2)$
  and thus vanishes in the  abelian  $U(1)$  case.}


One  may  view the  running of $\z$ as an RG flow in the effective 1d  defect theory coupled  to the bulk SYM theory.  This  interpretation   may be made more explicit by 
representing the path ordering in \rf{1.1}  using the  auxiliary  1d fermion  path integral as in 
\cite{Gervais:1979fv,Arefeva:1980zd,Brandt:1981kf,Hoyos:2018jky} 
and thus  getting an interacting 1d defect  action.  Considering  a  circular 
 contour, $F=-\log \W$  may be interpreted as a  1d  defect theory free energy on  $S^1$ (normalized by 
the partition function of the bulk theory). It is then natural to expect 
that this quantity provides a defect analog of the F-theorem. Specifically, the $d=1$ version of the generalized F-theorem   \cite{Klebanov:2011gs, Giombi:2014xxa, Fei:2015oha} adapted to defects \cite{Kobayashi:2018lil} requires that $  \widetilde F \equiv  {\te \sin { \pi d \ov 2} }\, \log Z({S^d}) \Big|_{d=1}  = \log Z({S^1}) =-F=\log \W$ decreases under RG flow: $\tilde F_{\rm UV}>\tilde F_{\rm IR}$. This is analogous to the  $g$-theorem \ci{Affleck:1991tk,Friedan:2003yc} that  applies  to a 1d boundary of a 2d theory.  The beta-function (\ref{1.3}) implies that $\z=0$ is the UV fixed point and 
$\z= 1$ the IR one, and so one shall find that
\begin{equation}
\log \vev{W^{(\z=0)}} > \log \vev{W^{(\z=1)}}\,. 
\end{equation}
This was indeed verified in \cite{Beccaria:2017rbe} to hold both in perturbation theory and in the strong coupling expansion.  
A proof of the  $d=1$ defect  version of the F-theorem was recently proposed in \cite{Cuomo:2021rkm}, where a quantity that monotonically decreases along the flow (and coincides with $\tilde F$ at fixed points) was also given.

On general   grounds, consistent with the interpretation of $\log   \W $ as a defect free enerenergygy,  we should have 
 \be \la{1.4}
 {\del \ov \del \z } \log   \W  =   \C \, \beta_\z  \ ,
 \ee
 where  the function 
 $\C=\C(\l,\z)$   has  the weak  coupling expansion $\C= {\l\ov 4}  +   \OO(\l^2)$ \ci{Beccaria:2017rbe}.
 \iffa 
  is expected to have no zeroes, since  $\z=1$  and $\z=0$   are  the only extrema  (minimum  and maximum)   of 
 $\vev{\WZ}$. \footnote{Relation (\ref{1.4}) follows from interpreting  $\vev{ \WZ }$ as a 1d  partition function  on $S^1$
 and doing conformal perturbation theory   near the $\z=1$  or $\z=0$    conformal points. Then, (\ref{1.4})
 is the one-dimensional case of the general relation   
   ${\del F\ov  \del g_i}  = \C^{ij} \beta_j $    for the free   energy $F$ on a sphere $S^d$ 
  computed by perturbing a CFT$_d$  by  a  linear combination of  the operators $g_i  O^i$
   (see, e.g.,  \ci{Klebanov:2011gs,Fei:2015oha}). }
   \fi
   
   In the case of a circle or straight line the flow of $\z(\mu)$ 
 is driven by the scalar operator $\phi_{6}\equiv \phi (x(\tau))$  in \rf{1.1}   restricted to the line. 
 Then  ${\del \ov \del \z } \vev{ \WZ } \big|_{\z=0, 1}  =0$  implies that its 
one-point function vanishes at the fixed  points, as required  by the 1d conformal invariance on the defect line.
From  \rf{1.4}  we also have 
 \be \la{1.5}
 {\del^2 \ov \del \z^2 }\log \W \Big|_{\z=0, 1}  =   \C\,  {\del \beta_{\z}  \ov \del \z} \Big|_{\z=0, 1} 
 \ . \ee
According to \rf{1.1} this    second derivative  is  
  given by the integrated  2-point function  of $\phi$ restricted to the line 
  \ci{Beccaria:2017rbe}.  The latter is  
 determined   by the  corresponding anomalous   dimensions at $\z=0$ and $\z=1$.
 Indeed,    from \rf{1.3} one finds that  
 $ {\del \beta_{\z}  \ov \del \z} $
  reproduces  \ci{Polchinski:2011im} the  leading  weak-coupling  terms in the anomalous dimensions \ci{Alday:2007he}    of $\phi$ at   the conformal points $\z=1$ and $\z=0$
  \ba
\la{1.6}  
 &\Dze  -1=  {\del \beta_\z  \ov \del \z} 
 \ ,  \\ 
 &\Do = 1 + {\l \ov  4 \pi^2}  + \OO(\l^2) \ , \qquad \ \ \ \ \  \ \  \Dz   = 1 - {\l \ov  8 \pi^2}  + \OO(\l^2)  \ . \la{16}
 \ea
 
 Our  main aim here    will be to find  the   $\l^2$ term in  the beta-function \rf{1.3}  which  should also  exhibit the factor $\z (1-\z^2)$.
Let   us start with writing down the general structure of $\b_\z$
\ba
\la{1.7}
 \beta_{\zeta} = b_{1}\,\l\,\z(1-\z^{2})+\l^{2}\,\z\,(1-\z^{2})\,(b_{2}+b_{3}\,\z^{2})+\l^{3}\,\z\,(1-\z^{2})\,
(b_{4}+b_{5}\,\z^{2}+b_{6}\,\z^{4})+\mc O(\l^{4})\ , 
\ea
where $b_{1} = -\tfrac{1}{8\pi^{2}}$ (cf. \rf{1.3}) 
 and $b_2,b_3, ...$ are to be determined. 
The dependence on powers of $\z$  at each order in $\l$  follows  from the  structure of the relevant perturbation-theory diagrams. 

An important  observation is that the coefficients of the 
highest $\z^{2n+1}$ powers  at  each $\l^n$  order in \rf{1.7}, \ie $b_{1}, b_{3}, b_{6}, \dots$, 
are determined by diagrams  with maximal  number of scalar  propagators  attached to the line. 
Thus they should not  
have  internal vertices, \ie  should be  given  just   by the scalar ladders. 
We shall   compute  them    using  the  vertex renormalization method
of \ci{Dotsenko:1979wb}  generalizing the  one-loop  computation  in \ci{Polchinski:2011im}.
The resulting terms  in $\beta_{\z}$  may be written as 
\ba
& \beta_{\z}^{\rm ladder} =
q_{1}\,\frac{\lambda}{4\pi^2}\,\z^{3}
+q_{2}\,\Big(\frac{\lambda}{4\pi^2}\Big)^{2}\,\z^{5}
+q_{3}\,\Big(\frac{\lambda}{4\pi^2}\Big)^{3}\,\z^{7}
+q_{4}\,\Big(\frac{\lambda}{4\pi^2}\Big)^{4}\,\z^{9}
+q_{5}\,\Big(\frac{\lambda}{4\pi^2}\Big)^{5}\,\z^{11}
+\cdots\ , \no   \\
& 
q_{1} = \frac{1}{2},\qquad
q_{2} = -\frac{1}{4},\qquad
q_{3} = \frac{1}{4}-\frac{\zeta_{2}}{8}, \qquad
q_{4} = -\frac{17}{48}+\frac{\z_{2}}{3}-\frac{\z_{3}}{12}, \la{1.12} \\
& 
q_{5} = \frac{29}{48}-\frac{37\,\z_{2}}{48}+\frac{29\,\z_{3}}{96}+\frac{25\,\z_{4}}{128}, 
\qquad \qquad   b_{n(n+1)\ov 2}= - {1\ov (4 \pi^2)^n} q_n \ , \no
\ea
where ${\zeta}_{n}\equiv \zeta(n)$ are the Riemann zeta-function values.
In particular, at the  two-loop order, we  get   from $q_2$ 
\be b_{3}=\frac{1}{4}\frac{1}{(4\pi^{2})^{2}}  \ . \la{b63} \ee
To find the five-loop  expression  for the $\beta_{\z}^{\rm ladder} $ in \rf{1.12}   we used  planar loop equation \rf{2.5}   and  dimensional regularization. As the  higher  coefficients  $q_n$  are transcendental, the exact  expression  for this planar  ladder-theory beta-function should be complicated.  Note that as  this  is  essentially a one-coupling  ($\xi\equiv  \l \z^2$)  model,  
  the  three  and higher loop coefficients  are scheme-dependent (see below).

An indirect way to fix   some  combinations of  other coefficients in the
 full  beta-function 
\rf{1.7} 
 is to use the relation  \rf{1.6}  between $\beta_\z$ and the anomalous dimension of the scalar operator 
  and the  value of  $\Do$ that  was  already  found earlier  
  from  diagrammatic computation for a cusp line in  \cite{Bruser:2018jnc}  (at the  two-loop level)  
and   from  the 
 quantum spectral curve  in \cite{Grabner:2020nis} (at several higher loop orders). 
 Explicitly, according to  \cite{Grabner:2020nis}
 \ba
&\qquad \qquad  \Do-1 = 
d_{1}\, \frac{\lambda}{4\pi^2}
+d_{2}\,\Big(\frac{\lambda}{4\pi^2}\Big)^{2}
+d_{3}\,\Big(\frac{\lambda}{4\pi^2}\Big)^{3} 
+d_{4}\, \Big(\frac{\lambda}{4\pi^2}\Big)^{4}
+\cdots\ , \la{113} \\
& d_{1} = 1, \qquad
d_{2} = -1, \qquad
d_{3} = 2-\frac{7\, \z_{4}}{4}, \qquad
d_{4} = -5+\zeta_{2}+\frac{\z_{3}}{2} -\frac{ \zeta_{3} \z_{2}}{2}-\frac{5\, \z_{5}}{8} +\frac{119\,\zeta_{6}}{16}, ... \la{1.14}
\ea
Comparing this with  \rf{1.6},\rf{1.7}
gives 
\ba
& b_1= -  {d_1\ov 2 (4 \pi^2)}  \ , \qquad \qquad \qquad \qquad \ \ 
b_{2}+b_{3}= -  {d_2\ov 2(4 \pi^2)^2 }  \ , \quad \la{17} \\
&b_{4}+b_{5}+b_{6}  = -   {d_3\ov2 (4 \pi^2)^3 }  \ , \qquad \qquad 
b_{7}+b_8+b_9 + b_{10} = -   {d_4\ov2 (4 \pi^2)^4 }  \ , ...
\la{1.17}
\ea
Using  \rf{b63}    then \rf{17} implies that 
\be \la{two}
b_{2} = \frac{1}{4}{1\ov (4\pi^{2})^2} \ .
\ee
Thus  the  explicit  form of the 2-loop $\beta_\z$ is given by 
\be
\la{1.8}
\beta_{\z} = -\frac{\lambda}{8\pi^{2}}\,\z\,(1-\z^{2})+
\frac{\lambda^{2}}{64\pi^{4}}\,\z(1-\z^{4})+\mc O(\lambda^{3})  \ . 
\ee
This in turn implies that  the two-loop terms in the  anomalous   dimensions \rf{16}  are given by 
\be \la{1.10}
\Do = 1+\frac{\lambda}{4\pi^2}-\frac{\lambda^{2}}{16\pi^4}+\mc O(\lambda^{3}) , \qquad\qquad
\Dz  = 1-\frac{\lambda}{8\pi^2}+\frac{\lambda^{2}}{64\pi^4}+\mc O(\lambda^{3}) \ , 
\ee
where $\Do$  is of course the same as in \rf{1.14}  while  the two-loop term in 
$\Dz $  is a new non-trivial result.\foot{It would    be interesting to reproduce
 it  by a direct  diagrammatic approach 
  similar to the one in   \cite{Bruser:2018jnc}.}
 Note that the sign-alternating   structure of  $\l$-expansion in  \rf{1.8} and \rf{1.10} 
is consistent with expectation that the planar  weak coupling expansion should have a 
finite radius  of convergence $|\frac{\lambda}{4\pi^{2}}|=1$ with the  strong-coupling $\l \gg 1$  asymptotics 
\ci{Alday:2007he,Giombi:2017cqn,
Beccaria:2017rbe}
$\Do = 2-\frac{5}{\sql}+\mc O({1\ov \l}) , \ \ 
\Dz =\frac{5}{\sql}+\mc O({1\ov \l}). $
\foot{Ref.   \cite{Grabner:2020nis} 
found also several   higher-order terms  in the strong-coupling expansion, correcting the leading terms \ci{Giombi:2017cqn} in 
 $\Do = 2-\frac{5}{\sql}+ ...$.
 These corrections were also obtained analytically  from the bootstrap approach in 
 \ci{Ferrero:2021bsb}.}

It is important to  stress that 
the two-loop  coefficients $b_{2}, b_{3}$ in   \rf{1.7},(\ref{1.8}),  are scheme independent:
 they are invariant  under 
 redefinitions of $\z$ that do not change the positions of the fixed points 
\be
\la{1.18}
\z' = \z+\z\,(1-\z^{2})\,\big[\l\, z_{1}+\l^{2}\,(z_{2}+z_{3}\,\z^{2})+\cdots\big] \ . 
\ee
Since the $\b_\z$ transforms  as a vector, we find that the coefficients in \rf{1.7}   change as  
\ba
b'_{i}=b_{i}+\delta b_{i} \ , \qquad  
 \delta b_{1}=\delta b_{2}=\delta b_{3}=\delta b_{4}=0, \qquad
 \delta b_{5} = -\delta b_{6} = b_{1}\,(3z_{1}^{2}+2z_{3})-2 b_{3}z_{1}.
\ea
This means that the beta-function is scheme independent at two loops, 
while  the invariant combinations  of the  three-loop  coefficients are $b_{4}$ 
and $b_{5}+b_{6}$.
This is  consistent with the fact that the dimensions
$\Dz = 1+   b_1 \l + b_2 \l^2 + b_4 \l^3 + ...$    and 
 $\Do$ in (\ref{1.14})  (and thus $d_n$ in \rf{17},\rf{1.17})
   should be scheme independent.


Let us   now  comment on  the  implications of the  above discussion for the 
  structure of higher order terms  the  expectation value of 
$\vev{\WZ}$  in \rf{1.2} on a circle (see  \ci{Beccaria:2017rbe,Beccaria:2018ocq}).
Since  the first derivative of $\W=\vev{\WZ}$   at the conformal points $\z=0,1$  should   vanish (cf. \rf{1.4}), we should    have 
\be
\la{1.21}
\W = \vev{W^{(1)}}\,\Big[1+w_{1}\,\l^{2}\,(1-\z^{2})^{2}+\l^{3}\,(1-\z^{2})^{2}(w_{2}+w_{3}\,\z^{2})+\cdots\Big]\ , \ee
where 
\cite{Erickson:2000af,Drukker:2000rr,Pestun:2007rz} 
\be
\la{1.24}
\vev{W^{(1)}} = \frac{2}{\sqrt\lambda}\,I_{1}(\sqrt\lambda) = 1+\frac{\lambda}{8}+\frac{\lambda^{2}}{192}+\mc O(\lambda^{3})\ .
 \ee
The coefficients 
$w_1$  and $w_2$ are scheme-independent, while $w_3 $ is  finite after renormalization of $\z$  and  in general  contains $\log \mu$   dependence  on the renormalization scale \ci{Beccaria:2018ocq}.  Using the general relation \rf{1.4} to the  beta-function  \rf{1.7} where 
 on general grounds   we should have 
 \be
\la{h1}
\mc C = \l\, c_{1} + \lambda^{2}(c_{2}+c_{3}\z^{2})+\mc O(\l^3) \ , 
\ee
we find  
\ba
\la{1.22}
w_{1} = -\frac{1}{4} b_{1}c_{1}, \qquad 
w_{2} = -\frac{1}{12}[b_{1}\,(3c_{2}+c_{3})+(3b_{2}+b_{3})c_{1}], \qquad
w_{3} = -\frac{1}{6}(b_{1}c_{3}+b_{3}c_{1})\ .
\ea
As   $b_{1}, b_{2}, b_{3}, w_2$ are scheme
independent while $w_3$ is not  this implies that $c_{3}$ in (\ref{h1}) is scheme dependent but 
$3c_{2}+c_{3}$ is scheme independent.\footnote{This    of course 
 follows also directly from the transformation law $\mc C'(\z') = (\frac{d\z'}{dz})^{-2}\,\mc C(\z)$  with  $\z'$  in 
(\ref{1.18}):    $ c'_{2}=c_2 -2c_{1}z_{1}$ and $c'_3= c_{3}+ 6 c_{1}z_{1}$ so that $3c'_{2}+c'_{3}= 3 c_2 + c_3$.
In general,  for a set of  couplings  $\z^i$  we have 
$\beta^i = \mu\,d\z^i/d\mu$   transforming as  a vector under their redefinitions and 
the relation 
$\frac{\partial}{\partial\z^{i}}\log \W   = \mc C_{ij}  \beta^j$
means that $\mc C_{ij}$  transforms as  a tensor. At a conformal point  $\beta^i(\z^{*}) =0$, one has
$ \frac{\partial^{2}}{\partial\z^{i}\partial\z^{j}} \log {\W}|_{\z=\z^{*}} =  \mc C_{ij}\,\frac{\partial\beta^{j}}{\partial\z^{i}}|_{\z=\z^{*}}$.
Here 
${\partial\beta^j\ov \partial\z^{i}}|_{\z=\z^{*}}$   does not  transform under 
$\z'= \z + X(\z)$ that does not change the position 
of the conformal point,  \ie  if $X(\z^{*})=0$. 
}
Here   \ci{Beccaria:2017rbe} 
\be 
 \la{1.26}
 w_{1} = \frac{1}{128\pi^{2}} \ ,\qquad \qquad    c_1= {1\ov 4}\ .
 \ee
 $w_{3}$  is determined   by the  contribution of   scalar 
  ladder diagrams (which is not UV  finite at order $\lambda^{3}$). Its renormalized value    was 
found   (using a particular 
 regularization scheme)  in \cite{Beccaria:2018ocq}  to be  
  $w_{3} =- \frac{1}{96\,(4\pi^{2})^{2}}\big[5+6\log(\mu R)\big]$. Here $R$ is the radius of the circle  and
   the coefficient  6 of $\log \mu$  in the bracket  is  related to 
 the coefficient in the  one-loop   beta-function \rf{1.3} (cf. \rf{1.2})
   while 5 is  scheme-dependent. In view of \rf{1.22} this then also  fixes the value of  $c_3$  in the same scheme.
The   value of  $w_2$   remains currently unknown. 
 If we   choose a scheme   in which the value of $w_3$   is  equal to $- w_2$
 then the $\l^3$  term in \rf{1.21}   will   be proportional to $(1-\z^2)^3$. 
One can then conjecture that   there   exists a scheme in which similar   simplification 
  happens to all orders in $\l$
  \cite{Beccaria:2017rbe}, \ie   one gets 
\be
\la{1.29}
\vev{\WZ} = \vev{W^{(1)}}\,\Big[1+\mc F\big(\l\,(1-\z^{2})\big)\Big],
\ee
where  the function $\mc F(x)$  has a regular power series expansion 
$\mc F(x) = w_{1}\,x+w_{2}\,x^{2}+\cdots$. 

\

The rest   of this paper is organized   as follows. 
In section 2    we shall   explain how to compute the  coefficients in the  beta-function \rf{1.12}  in ladder  approximation  using dimensional regularization. We shall  mention that defining the effective ladder theory in $d=4-\eps$  dimensions the
corresponding   beta-function has a Wilson-Fisher-like  zero  and the  
 value of the  Wilson loop at the corresponding  IR fixed point  is consistent   with the   1d defect version of F-theorem  \cite{Beccaria:2017rbe,Cuomo:2021rkm}.
 
 In section 3  we shall consider the computation of  two-point functions of scalars 
 inserted on the Wilson loop at two loops in ladder approximation. We shall consider  separately the cases of a  ``transverse'' scalar  not coupled to the loop  and the  scalar coupled to the loop. Using the  Callan-Symanzik equation we shall  find the corresponding   
 anomalous dimensions 
of   the   scalar operators  
 relating them to the beta-function $\b_\z$  and its derivative. 
 
There  are  also a few  technical Appendices.  In particular, in Appendix \ref{PS}
we shall  present some details of the  computation of the linear in $\z$ term in $\b_\z$ 
in \rf{1.3}  which were not spelled out in \ci{Polchinski:2011im}.


\section{Scalar ladder  contributions to the beta-function}

 In this section we shall present the derivation of  the  coefficients of the $\l^n \z^{2n+1} $  terms   in  the beta-function \rf{1.12}.

\subsection{Vertex renormalization method}

\iffa
that  idea of looking at  2-point of transverse scalar  may be simpler  indeed 
  that trying   via   one-point function as  at 1 loop   here  in  diagrams  in 
your  1.16   there is  only 2  integrals in 1.16 while in 1-point function of coupled   scalar
 computed the way   I was suggesting  there will     be 3  integrals   already at leading order. 
this   seems important point to stress  later.  This  simplification was absent in pure YM case.
Basically,  either we stay with W  and compute  it  high enough   order  to see UV divergences and thus 
 how z  should be renormalized  or   we introduce extra probes --  1-point or 2-point with un-integrated  points 
and that allows  us to decompactify the circle to line  to study UV  singularities from  integrated points coming close to each   other. 
\fi

To compute the highest in $\z$ terms in each order of the small $\l$ expansion in the beta-function 
  we need to consider   ladder  graphs with  only  scalar 
  propagators attached to Wilson loop. 
The perturbative expansion of the corresponding 
 $\vev{\WZ}^{\rm ladder}$  is  then   given in the planar limit 
 by  a  power series in a single  effective  coupling  $\xi$
\be \vev{\WZ}^{\rm ladder} = W(\xi),  \qquad \qquad 
\xi\equiv \lambda\,\z^{2} \ , \la{21}
\ee
 where  (for a closed contour parameterized by $\tau\in (0, 2 \pi)$)  
\begin{align}
\la{2.3}
W = \GW (2\pi),\qquad  \qquad \GW (\tau) =\lim_{N\to \infty}
{ 1 \over N}  \Big\langle \text{Tr}\,\text{P}\,\exp\int_0^\tau
d\tau'\, \phi(\tau')\Big\rangle \ . 
\end{align}
Compared to \rf{1.1} 
 we have redefined the scalar $\phi\to \z^{-1}\phi$, and  set 
 $\phi(\tau) \equiv \phi(x(\tau))$. The averaging is  done 
in the free adjoint scalar theory with the action  
 \be \la{act}
 S = \frac{N}{\xi}\,\int d^4 x\  \text{Tr} (\partial \phi \partial \phi)  \ . \ee
 In what follows we shall  consider 
 the case of a circular or straight line contour when the (unregularized) 
 propagator $D(\tau-\tau') = \langle  \phi(\tau)\,\phi(\tau')\rangle$  has the following form\footnote{\la{f2} 
The original  $\mc N=4$ SYM action is schematically $S= \frac{1}{g^{2}_{{\text{YM}}}}\,\int d^4 x\  \text{Tr} (F^{2}+D \phi D \phi+\phi^{4}+\dots)$,
and $\l=g^{2}_{{\text{YM}}}\,N$.
Eq. (\ref{2.4}) takes into account a factor $1/2$ 
from $T^{a}T^{a} = \frac{N}{2}\bm{1}$, valid for the generators $T^a$ of $SU(N)$ in the fundamental representation.}
\be
\la{2.4}
\text{circle:}\quad D(\tau) =  \frac{\xi}{8\pi^{2}}\frac{1}{4\,\sin^{2}\frac{\tau}{2}}, \qquad\qquad 
\text{line:}\quad D(\tau) =  \frac{\xi}{8\pi^{2}}\frac{1}{\tau^{2}}.
\ee
The problem of computing  (\ref{2.3})  with  (\ref{2.4}) is well defined   and 
  we expect $W$ to admit a renormalizable perturbative 
expansion in $\xi$ such that all UV divergences can   be  absorbed into a 
 redefinition of $\xi$ (any  possible 
multiplicative renormalization of the loop operator  should be  absent 
  in dimensional regularization).

In the planar limit
the function $\GW (\tau)$ in \rf{2.3} 
obeys the  following integral equation\footnote{K. Zarembo, private communication.}
\begin{align}
\la{2.5}
\frac{\partial\GW (\tau)}{\partial \tau} = \int_0^\tau d\tau'\,\GW (\tau')\,\GW (\tau-\tau')\,D(\tau-\tau').
\end{align}
Note that  by definition in \rf{2.3}  we have 
 $\GW (0)=1$  and  in \rf{2.5}  one has also $  \GW '(0) =0$. Eq.~(2.5) follows upon differentiation of the integral equation that  is implied   by 
 the structure of the planar expansion of (\ref{2.3}) 
 in the free  scalar theory (here thick line denotes  the contour parameterised by $\tau$)
\ba
\begin{tikzpicture}[line width=1 pt, scale=0.5, baseline=0]
\node[left] at (0,0) {$0$};
\node[right] at (4,0) {$\tau$};
\draw[line width=2 pt] (0,0)--(4,0);
\draw[fill=lightgray] (2,0) circle(1);
\node at (2,-1.75) {$\GW (\tau)$};
\end{tikzpicture} = \int_{0}^{\tau}d\tau'\,\int_{0}^{\tau'}d\tau''\ 
\begin{tikzpicture}[line width=1 pt, scale=0.5, baseline=0]
\node[left] at (0,0) {$0$};
\node[right] at (10,0) {$\tau$};
\draw[line width=2 pt] (0,0)--(10,0);
\draw[fill=lightgray] (2,0) circle(1);
\node at (2,-1.75) {$\GW (\tau'')$};
\draw[fill=lightgray] (6,0) circle(1);
\node at (6,-1.75) {$\GW (\tau'-\tau'')$};
\node[below] at (4,0) {$\tau''$};
\node[below] at (8,0) {$\tau'$};
\draw (4,0) arc(180:0:2);
\node at (6,2.4) {$D(\tau''-\tau')$};
\end{tikzpicture}
\la{25}
\ea
Eq. (\ref{2.5}) is valid for any propagator $D$.
 In particular, it can be  applied to the case of the ladder  contributions to 
 the  expectation value  of WML (i.e. $\z=1$ case of \rf{1.1})  for a circle  when the  effective propagator  is constant \cite{Erickson:2000af}
\be\Big\langle {[iA^{a}(\tau)+\phi^{a}(\tau)]\,[iA^{b}(\tau')+\phi^{b}(\tau')]} \Big\rangle= \delta^{ab}\,\frac{\l}{8\pi^{2}\,N}, \qquad \qquad 
\la{2.7}
D(\tau) =  D_0 = \frac{\lambda}{16\,\pi^2} , 
\ee
where  we used that  $T^{a}T^{a}=\frac{N}{2}\bm{1}$.
Taking  the Laplace transform of (\ref{2.5}) we have $-1+s\,\widetilde\GW (s) =  D_{0}\,\widetilde{\GW }^{2}(s)$  and 
 thus finally 
\be
\widetilde\GW (s) = \frac{s}{2 D_{0}}\,\Big(1-\sqrt{1-\frac{4 D_{0}}{s^2}}\Big)\ \ \ \to \ \ \ 
\GW (\tau) = \frac{1}{\tau\sqrt{D_{0}}}\,{I}_1(2\,\tau\,\sqrt{D_{0}}),
\ee
reproducing  (\ref{1.24}) after setting $\tau=2 \pi$ as in \rf{2.3}.  Similar Dyson integral equations also 
appear for the correlation function of two Wilson loops \cite{Correa:2018pfn}. 

For a non-constant propagator $D(\tau)$, the solution of the loop equation  \rf{2.5} appears to 
be highly  non-trivial, in particular,  due to divergences starting at three loops. 
In Appendix  ~\ref{A}  we demonstrate  how to use \rf{2.5} 
  to reproduce   the two-loop result  for \rf{1.21}, i.e. find the value of $w_1$ in (\ref{1.26}).

Following \cite{Dotsenko:1979wb} and adapting their discussion to the present case, to compute the renormalization of $\z$ it is useful to consider  the ``one-point''  scalar correlator 
on  a   straight 
 Wilson line segment. Before performing the rescaling of $\zeta$ into $\phi$ introduced above, one starts with the quantity
\begin{equation} 
{ \langle  \tr\Big(\phi(\tau_{0})\,\text{P}\exp\big[ 
\int_{\tau_{1}}^{\tau_{2}}d\tau\,\zeta \phi(\tau)\big] \Big)\rangle \over   
\langle  \tr\Big(\text{P}\exp \big[ 
\int_{\tau_{1}}^{\tau_{2}}d\tau\,\zeta \phi(\tau)\big] \Big)\rangle} \,.
\label{vert-ren-start}
\end{equation}
Note that this is not exactly the standard one-point function of the operator $\phi$ on the Wilson line, because we are integrating only over a segment $\tau_1<\tau<\tau_2$
(with far-separated points $\tau_1, \tau_2$) 
 and $\tau_0$  is also  assumed 
to be far  from that interval (hence $\phi(\tau_0)$ does not participate in the path-ordering).  Following   \cite{Dotsenko:1979wb},   all UV divergences    should come from   ``internal''
coinciding points  not involving $\tau_0, \tau_1, \tau_2$, so 
it is sufficient to look at this object in order 
to find the renormalization of $\z$ (or $\xi$). The normalization in (\ref{vert-ren-start}) by the expectation value of the Wilson line segment without insertion is needed 
in order to remove some spurious divergences associated with the finite endpoints \cite{Dotsenko:1979wb}.  After performing the rescaling $\phi\to \z^{-1}\phi$, eq.~(\ref{vert-ren-start}) 
becomes
\begin{equation} 
{ \langle  \tr\Big(\phi(\tau_{0})\,\text{P}\exp\big[ 
\int_{\tau_{1}}^{\tau_{2}}d\tau\, \phi(\tau)\big] \Big)\rangle \over   \zeta\,
\langle  \tr\Big(\text{P}\exp \big[ 
\int_{\tau_{1}}^{\tau_{2}}d\tau\, \phi(\tau)\big] \Big)\rangle} .
\label{vert-ren-step2}
\end{equation}
In the planar limit, the numerator in this expression satisfies  a relation analogous to \rf{25} and can be written as
\ba
\la{2.10}
\lim_{N\to\infty}\frac{1}{N}\langle & \tr\Big[\phi(\tau_{0})\,\text{P}\exp
\int_{\tau_{1}}^{\tau_{2}}d\tau\, \phi(\tau) \Big]\rangle 
=  \int_{\tau_{1}}^{\tau_{2}} d\tau\, \GW (\tau-\tau_{1})\,D(\tau_{0}-\tau)\,\GW (\tau_{2}-\tau).
\ea
On the other hand, the denominator factor in (\ref{vert-ren-step2}) is simply equal to $\zeta \GW(\tau_2-\tau_1)$. When performing the renormalization of (\ref{vert-ren-step2}), we 
may effectively factor out the $\tau$ integration on the right-hand-side of (\ref{2.10}),  since the integral over $\tau$ cannot  bring in new UV divergences as the point $\tau_0$  is supposed to be 
far away from $\tau_1, \tau_2$. Therefore, one can see that in the planar limit the calculation reduces to study the renormalization of the ``vertex function''
\be\la{29} 
V= \xi\, \GW (\tau_{1})\,\GW (\tau_{2})\ ,
\ee
where $\tau_{1}, \tau_{2}$ are  arbitrary (far separated)  fixed points on the line. Note that at planar level the normalization by the denominator factor in (\ref{vert-ren-step2}) does not play an important role since, being 
equal to $\zeta \GW(\tau_2-\tau_1)$, it is essentially the same as the ``square root'' of the $V$ function defined above (recall  that $\xi=\lambda \zeta^2$), and hence it is finite once $V$ is made finite by renormalization (or vice-versa). At the non-planar level, however, the denominator factor is expected to play a non-trivial role. 
To summarize, in the planar limit  it will be sufficient  
  to consider  the renormalization of $V$ as given in \rf{29}. 
The relevant  collection of diagrams  may be represented symbolically as 
\be
\la{2.12}
\begin{tikzpicture}[line width=1 pt, scale=0.5, baseline=0]
\node[above] at (0,2) {$\phi(x_{0})$}; 
\node[below] at (0,0) {$0$};
\node[right] at (6,0) {$\tau_{1}$};
\node[left] at (-6,0) {$\tau_{2}$};
\draw (0,2)--(0,0);
\draw[line width=2 pt] (-6,0)--(6,0);
\draw[fill=lightgray] (3,0) circle(1.5);
\node[above] at (3,2) {$\GW (\tau_{1})$};
\draw[fill=lightgray] (-3,0) circle(1.5);
\node[above] at (-3,2) {$\GW (\tau_{2})$};
\end{tikzpicture}
\ee
Here  we  have  chosen  the ``middle''  point $\tau$  in \rf{2.10}  as  0  (using translational invariance)   and  replaced  the argument $\tau_0$  of $\phi$ by a  generic point $x_0$ that may or may not lie on the line. The  vertical line  represents  the propagator $D(\tau_0-\tau)$ or $D(x_{0}-x(0))$  which  will play only a spectator role. 

 The  strategy   is   to  find  the  divergent part of  $V$ in  \rf{29}
and then absorb  the divergences into the renormalization of $\xi$. This requires   computing $\GW $  from the corresponding sum of  planar  ladder diagrams   (or using the loop  equation \rf{2.5}).

Let us  note that the reason why   this  ``scalar ladder''  model  is effectively an  interacting  one  (despite  the bulk theory being free)  is due to the path ordering in \rf{2.3}.  As was 
 mentioned in the Introduction, one can   cast the problem of  
 renormalization of $\z$ or  $\xi$ in a more standard form     by  representing the path ordering  using a functional   integral over 1d fermions $\chi^i(\tau)$ in the fundamental representation. Integrating  first over the  free adjoint  bulk 
 scalar field   we  then get an effective  1d action of the following  schematic form
\be\la{211}
I \sim  \int d\tau\,\bar\chi_i \dot\chi^i +\xi  \int d\tau\,d\tau'\,  \bar\chi_j(\tau) \chi^i(\tau)\,  \frac{1}{|\tau-\tau'|^2}\, \bar\chi_i(\tau') \chi^j(\tau')\ .
\ee
Introducing a cutoff into the propagator  and  expanding  in $\xi$ one should be able then to  compute the corresponding $\beta_\xi$ in the usual  way.


\subsection{Five-loop beta-function in ladder approximation}

Let us consider the loop equation (\ref{2.5}) on a line with a simple analytic   regularization
of the propagator
\be
\la{2.14}
D(\tau-\tau') = \frac{\xi}{8\pi^{2}}\frac{1}{|\tau-\tau'|^{2-\eps}}.
\ee
This is essentially the standard   dimensional regularization with $d=4-\eps$, $\eps>0$,  where we did not
include the usual   $\eps$-dependent normalization factor
 (this is equivalent to a redefinition of the renormalization scale).
Solving the loop equation perturbatively, \ie expanding in powers of $\xi$ 
\be
\la{2.15}
\GW (\tau) =1 +  \xi\,\GW _{1}(\tau)+\xi^{2}\,\GW _{2}(\tau)+\cdots\ ,
\ee
we find   at the leading order
\be
\frac{\partial}{\partial\tau}\GW _{1}(\tau) = \frac{1}{\xi} \int_{0}^{\tau} d\tau' \, D(\tau-\tau') = \int_{0}^{\tau} d\tau' \frac{1}{8\pi^{2} (\tau-\tau')^{2-\eps}}
= -\frac{\tau^{-1+\eps}}{8\pi^{2} (1-\eps)}.
\ee
Integrating this with the boundary condition $\GW _{1}(0)=0$ gives 
\be
\GW _{1}(\tau) =-  \frac{\tau^{\eps }}{8 \pi ^2 (1-\eps ) \eps } \ . 
\ee
The corresponding   one-loop correction to the vertex (\ref{2.12}) is  represented  by
$ 1  \times \GW _1 (\tau_1) + \GW _1 (\tau_2) \times 1 $, i.e.   by the following diagrams
\be\la{215}
\begin{tikzpicture}[line width=1 pt, scale=0.5, baseline=0]
\node[above] at (0,2) {$\phi(x_{0})$}; 
\node[below] at (0,0) {$0$};
\node[right] at (3,0) {$\tau_{1}$};
\node[left] at (-3,0) {$\tau_{2}$};
\draw (0,2)--(0,0);
\draw[line width=2 pt] (-3,0)--(3,0);
\draw[fill=lightgray] (0,0) circle(1);
\end{tikzpicture} = 
\begin{tikzpicture}[line width=1 pt, scale=0.5, baseline=0]
\node[above] at (0,2) {$\phi(x_{0})$}; 
\node[below] at (0,0) {$0$};
\node[right] at (3,0) {$\tau_{1}$};
\node[left] at (-3,0) {$\tau_{2}$};
\draw (0,2)--(0,0);
\draw[line width=2 pt] (-3,0)--(3,0);
\draw (0.5,0) arc(180:0:1);
\end{tikzpicture} +
\begin{tikzpicture}[line width=1 pt, scale=0.5, baseline=0]
\node[above] at (0,2) {$\phi(x_{0})$}; 
\node[below] at (0,0) {$0$};
\node[right] at (3,0) {$\tau_{1}$};
\node[left] at (-3,0) {$\tau_{2}$};
\draw (0,2)--(0,0);
\draw[line width=2 pt] (-3,0)--(3,0);
\draw (-2.5,0) arc(180:0:1);
\end{tikzpicture} +\cdots\ .
\ee
Next,    $\GW _{2}(\tau)$ is given  by 
\be
\GW _{2}(\tau) = 
\begin{tikzpicture}[line width=1 pt, scale=0.5, baseline=0]
\node[left] at (0,0) {$0$};
\node[right] at (7,0) {$\tau$};
\draw[line width=2 pt] (0,0)--(7,0);
\draw (1,0) arc (180:0:1);
\draw (4,0) arc (180:0:1);
\end{tikzpicture}\qquad+\qquad
\begin{tikzpicture}[line width=1 pt, scale=0.5, baseline=0]
\node[left] at (0,0) {$0$};
\node[right] at (6,0) {$\tau$};
\draw[line width=2 pt] (0,0)--(6,0);
\draw (2,0) arc (180:0:1);
\draw (1.5,0) arc (180:0:1.5);
\end{tikzpicture}\ ,
\ee
with the explicit expression being 
\be
\GW _{2}(\tau) = \frac{\tau^{2 \eps } \big[ 2 \eps  (-1+2 \eps ) \Gamma (-1+\eps ) \Gamma 
(1+\eps )+\Gamma (1+2 \eps )\big]}{128 \pi ^4 (-1+\eps ) \eps ^2 (-1+2 
\eps ) \Gamma (1+2 \eps )}.
\ee
All  higher order functions $\GW _{n}(\tau)$ may be  expressed as 
\be
\la{L.0}
\GW _{n}(\tau) =  \frac{K_{n}(\eps)}{\Gamma(1+n\eps)}\,\tau^{n\eps},
\ee
where the coefficients $K_{n}(\eps)$ are determined by the recurrence relation
\be
\la{L.3}
K_{n}(\eps) = \frac{1}{8\pi^{2}}\,\sum_{p=0}^{n-1}K_{p}(\eps)\,K_{n-1-p}(\eps)\,\frac{\Gamma(-1+(p+1)\eps)}{\Gamma(1+p\eps)}, \qquad K_{0}(\eps)=1.
\ee 
This follows from the Laplace transform $\mc L$  of the loop equation 
on the line that reads\footnote{
The perturbative solution takes the form  $\wt\GW _{n}(s) = \frac{K_{n}(\eps)}{s^{1+n\eps}}$, \cf (\ref{L.0}). Replacing this in the loop equation and
using $(-\partial_{s})^{\alpha}\frac{1}{s^{\beta}} = \frac{\Gamma(\alpha+\beta)}{\Gamma(\beta)}\frac{1}{s^{\alpha+\beta}}$ gives  
(\ref{L.3}).
}
\be 
s\,\wt\GW (s)-1 = \frac{\xi}{8\pi^{2}}\,\wt \GW (s)\,\mc L\big[\frac{\GW (\tau)}{\tau^{2-\eps}}\big] = \frac{\xi}{8\pi^{2}}\,\wt \GW (s)\,(-\partial_{s})^{\eps-2}\wt\GW (s).
\ee
In the following we shall present the  explicit results  to    five loop order, \ie  including  $\GW _{5}(\tau)$ in \rf{2.15}.

We can then   find  the divergences  in \rf{29}, \ie  coefficients of 
poles   in $\epsilon\to 0$ in 
\ba
\la{2.23}
V=\xi\,\GW (\tau_{1})\,\GW (\tau_{2}) = \xi+V_{2}\, \xi^{2}+V_{3}\, \xi^{3}+\cdots \ 
\ea
and  cancel them by replacing  the bare coupling $\xi$ in terms of the renormalized  one using the  familiar general relation\foot{Note that the factor $\mu^\eps$ is required to match the dimensions: defining the   theory in $d=4-\eps$   dimensions 
we  keep the dimension of the scalar field  to be 1 (as it is coupled to the  line
which still has dimension 1)  so that the bulk action ${N\ov \xi} \int d^{4-\eps} x\, { \rm Tr }( \del \phi \del \phi)$   should have $\xi$   with  the mass dimension $\eps$. 
Equivalently, this follows from the fact that  the propagator \rf{2.14}   should 
still have dimension 2.}
\ba
\la{2.24}
\xi &= \mu^{\eps}\Big[\xi(\mu)+\frac{p_{11}}{\eps}\,[\xi(\mu)]^{2}+\left(\frac{p_{21}}{\eps}+\frac{p_{22}}{\eps^{2}}\right)\,[\xi(\mu)]^{3}
+\left(\frac{p_{31}}{\eps}+\frac{p_{32}}{\eps^{2}}+\frac{p_{33}}{\eps^{3}}\right)\,[\xi(\mu)]^{4}\lp\qquad \qquad 
+\left(\frac{p_{41}}{\eps}+\frac{p_{42}}{\eps^{2}}+\frac{p_{43}}{\eps^{3}}+\frac{p_{44}}{\eps^{4}}\right)\,[\xi(\mu)]^{5}+\cdots\Big].
\ea
The condition that the bare coupling  does not depend  on $\mu$ implies
 various relations   between the  coefficients  in \rf{2.24}  
 and  the resulting beta-function  expressed in terms of the renormalized coupling is given by 
 (see (\ref{B.9}),\rf{B.10})
 \be \la{22}
 \beta_\xi= \mu \frac{d}{d\mu}\xi
   = p_{11}\xi^{2}+2\,p_{21}\xi^{3}+3\,p_{31}\xi^{4}+ 4\,p_{41}\xi^{5} + \cdots \ . \ee
 Requiring that the vertex (\ref{2.23}) expressed in terms of $\xi(\mu)$ is finite  gives
\be
\la{2.25}
p_{11} = \frac{1}{4\pi^{2}}, \quad \ p_{21} = -\frac{1}{64\pi^{4}}, \quad \  p_{31} = \frac{12-\pi^{2}}{4608\pi^{6}}, \quad \ 
p_{41} = \frac{-51+8\pi^{2}-12\zeta_{3}}{73728\pi^{8}}, \ ...
\ee
and thus
\iffa 
\ba\la{bee}
\beta_{\xi} =& \frac{1}{4\pi^{2}}\,\xi^{2}-\frac{1}{32\pi^{4}}\,\xi^{3}+\left(\frac{1}{128\pi^{6}}-\frac{1}{1536\pi^{4}}\right)\,\xi^{4}
+\left(-\frac{17}{6144\pi^{8}}+\frac{1}{2304\pi^{6}}-\frac{\zeta_{3}}{1536\pi^{8}}\right)\,\xi^{5}\lp
+\left(\frac{29}{24576\pi^{10}}-\frac{37}{147456\pi^{8}}+\frac{5}{1179648\pi^{6}}+\frac{29\zeta_{3}}{49152\pi^{10}}\right)\,\xi^{6}
+\cdots\ ,
\ea\fi 
\ba
\la{2.27}
\beta_{\xi} &
= \xi\,\Big[
\frac{\xi}{4\pi^2}-\frac{1}{2}\left(\frac{\xi}{4\pi^2}\right)^2+\left(\frac{1}{2}-\frac{\zeta_{2}}{4}\right) 
\left(\frac{\xi}{4\pi^2}\right)^3+\left(-\frac{17}{24}+\frac{2\zeta_{2}}{3}-\frac{\zeta_{3}}{6}\right) 
\left(\frac{\xi}{4\pi^2}\right)^4\lp\qquad 
+\left(\frac{29}{24}-\frac{37 \zeta_{2}}{24}+\frac{29 \zeta_{3}}{48}+\frac{25\zeta_{4}}{64}
\right) \left(\frac{\xi}{4\pi^2}\right)^5+\cdots
\Big] \ , 
\ea
where we added also  the five-loop  contribution 
and  $\z_n$ are the zeta-function values. 
Recalling that according to \rf{21} $\xi=\l \z^2$  where $\l$ is not running 
 we may  then read off the ladder  contribution to the beta-function of $\z$
\be
\la{2.28}
\beta_{\xi} =   \mu \frac{d}{d\mu}(\lambda\zeta^{2}) = 2\,\lambda\,\zeta\,\beta^{\rm ladder}_{\zeta}\ , 
\ee
reproducing \rf{1.12}.  Explicitly, at 
the  two-loop order 
\be
\la{2.29}
\beta^{\rm ladder}_{\zeta} = \frac{\lambda}{8\pi^{2}}\zeta^{3} 
-\frac{\lambda^{2}}{64\pi^{4}}\zeta^{5} +\cdots\ . 
\ee
Note  that higher loop terms in $\beta^{\rm ladder}_{\zeta} $   obtained 
in this way   have similar transcendental structure to higher loop terms in $\Do$ 
in \rf{113},\rf{1.14} found in \ci{Grabner:2020nis}. 

\def \lad {{\rm ladder}}

\subsection{Comment on Wilson-Fisher fixed point} 

Let us  note that  considering the theory in $d=4-\eps$ dimensions   and thus  keeping the order $\eps$ term in the beta-function \rf{2.27}, 
  we get from \rf{2.24}
\be
\la{2.33}
\beta_{\xi} = -\eps\,\xi+\frac{\xi^{2}}{4\pi^{2}}-\frac{\xi^{3}}{32\pi^{4}}+\cdots.
\ee
This  implies that the effective 1d  theory corresponding to the scalar ladder   approximation  has,  in addition to the trivial UV fixed point $\xi=0$,  
 a  non-trivial  IR  fixed point  
\be
\la{2.34}
\xi^*=4\pi^{2}\eps+2\pi^{2}\eps^{2}+\cdots\, .
\ee
The expectation value of 
$\vev{\WZ}$ for a circle  found in $d=4-\eps$  in ladder   approximation    may be written as \cite{Beccaria:2017rbe}\footnote{Note that in  \cite{Beccaria:2017rbe} 
we used  $d=4-2\eps$ while  here  $d=4-\eps$.}
\be
\vev{{\WZ}}^{\rm ladder}\equiv W(\xi) 
= 1-\frac{\eps}{16}\,\xi +\frac{1}{128\pi^{2}}\,\xi^{2}+\cdots\, .
\ee
Evaluated at the fixed point \rf{2.34}  this gives 
\be
\la{2.36}
W(\xi^{*}) = 1-\frac{\pi^{2}}{8}\,\eps^{2}+\cdots.
\ee
Thus $\log W (\xi^{*})    <  \log W(0) =0$  in agreement with the 1d  defect  version of  F-theorem.  On the other hand, note that directly in $d=4$, the beta-function in the ladder approximation
is positive in perturbation theory, which means that the coupling $\xi$ grows in the UV. From the point of view of the F-theorem, this is consistent with the fact that $\log W(\xi)=\frac{1}{128\pi^{2}}\,\xi^{2}+\cdots$
is positive in perturbation theory. 

\iffa
As a final remark, we point out that in the adjoint scalar ladder theory
in $d=4$, the beta-function is positive in perturbation theory, $\beta_{\xi} = +\frac{\xi^{2}}{4\pi^{2}}+\mc O(\xi^{3})$.
Thus, the renormalized $\xi$ increases towards the UV. This is consistent with the fact that in this reduced
theory $\vev{W} = 1+\frac{\xi^{2}}{128\pi^{2}}+\mc O(\xi^{3})$ has a positive correction. Instead, as we have seen,
the non-trivial fixed point $\xi^{*}$ in $d=4-\eps$ is in the IR (due to the $-\eps\xi$ term in the 
beta-function), and so we need the leading correction 
to $\vev{W}$ to be negative as in (\ref{2.36}).
The two loop result (\ref{1.23}) implies the inequality 
\be 
\la{2.30}
\log \vev{ W^{(0)} }\  >\  \log \vev{ W^{(1)} } \ . 
\ee
This relation can be understood by regarding  $\vev{\WZ} = Z_{S^1} $ as the  partition function of a 1d  QFT on $S^{1}$. Then, 
the F-theoremf \cite{Myers:2010xs,Klebanov:2011gs,Casini:2012ei, Giombi:2014xxa,Fei:2015oha} in $d=1$ (where it corresponds to 
the $g$-theorem \ci{Affleck:1991tk,Friedan:2003yc} applying to the boundary of a 2d theory) 
requires
\be  
\widetilde F_{_{\rm UV}} > \widetilde F_{_{\rm IR}}\,,
\qquad\qquad  \widetilde F \equiv  {\te \sin { \pi d \ov 2} }\, \log Z_{S^d} \Big|_{d=1}  = \log Z_{S^1} =-F\, ,
\ee 
consistent with (\ref{2.30}). We also remark that $\vev{\WZ}$ in (\ref{1.23}) decreases monotonically with $\zeta$ 
from the non-supersymmetric $\z=0$ to the supersymmetric $\z=1$ fixed point.  \footnote{
At higher orders, assuming a scheme obeying (\ref{1.29}), the $g$-theorem
requires (non-perturbatively) $\mc F(x)>0$.
}
Notice that the function $\log\vev{\WZ}$  is not in general supposed to 
decrease monotonically with $\z$ from $\z=0$ to $\z=1$.
Along the flow, the proposal of the recent paper \cite{Cuomo:2021rkm}
is that the quantity which decreases under the flow is $s = (1-R \frac{d}{dR}) \log g$, 
where $\log g = \log \vev{\WZ}$ in our case (in general, $\log g = \log(Z_{\rm defect CFT}/Z_{\rm CFT})$).
Since $R \frac{d}{dR} = \mu \frac{\partial}{\partial\mu}$, we also can write
$s = \log\vev{\WZ}+\beta\frac{d}{d\z}\log\vev{\WZ}$.
This $s=s(\z)$ should be the quantity that monotonically decreases. Indeed, as discussed in \cite{Cuomo:2021rkm},
\be
\la{2.32}
\mu\frac{\partial}{\partial\mu}s(\z) = -\beta_{\z}\frac{\partial s}{\partial \z} = -\frac{\l^{3}}{256\pi^{4}}\z^{2}(1-\z^{2})^{2}+\mc O(\l^{4}),
\ee
which is negative, at least in perturbation theory. 
\fi

\section{Two-point function  of scalars on the Wilson line}

\iffa 
\subsection{Defect CFT data from  loop expectation value} 

As discussed in \cite{Beccaria:2017rbe} it is possible to relate the derivatives of $\vev{\WZ}$ with respect to $\z$ at $\z=0, 1$  and
conformal data of the defect theory. This differentiation produces integrated correlators of the conformal
 field $\phi=\phi_{6}$  coupled to the loop.\footnote{For a  discussion 
 of insertions of  general composite operators on the $\z$-dependent Wilson loop  see \cite{Correa:2018fgz} where it is shown that the one-loop dilatation operator for $SO(6)$ scalar insertions 
is integrable at $\z=0$ and $\z=1$.}
In the case of insertions on the WML ($\z=1$) for a circle\footnote{
The defect correlator of gauge theory operators in the adjoint representation  restricted to the line $t=\tau$  is  given  by 
$\llangle   \mc O(\tau_1)  \mc O(\tau_2)  \rrangle  \equiv 
 \langle {\rm Tr}\,{\PP} \big[\mc O(\tau_1)O(\tau_2)\ e^{\int d\t\, (iA_t+ \phi_6)}\big]\rangle$. We assume the normalization  such that  $\llangle 1 \rrangle=1$.
}
\be 
\la{3.1}
\lla  \phi (\t_1)\,  \phi  (\t_2) \rra^{(1)} 
=  \frac{C^{(1)}_{2}(\l) }{| 2\sin\frac{\tau_{12}}{2}  |^{2\Do}} \ ,  \qquad  C^{(1)}_{2} = \frac{\l}{8 \pi^2}  + \mc O(\l^2) \ , \qquad 
\Do  =1  +  \frac{\l}{4 \pi^2}\,  + \mc O(\l^{2}),
\ee 
where the leading term in $C_{2}(\lambda)$ is determined  by  the free field theory limit 
(cf. \rf{2.4}). 
In the   case  of  the standard   circular  WL   (with no scalar coupling, i.e. $\z=0$)   
 the defect  CFT$_1$   has unbroken $SO(6)$  symmetry  and all 6 scalars $\phi_{n}$
 have the same  correlator
 \be
  \llangle  \phi_{n}(\tau_1)  \phi_{m}(\tau_2)  \rrangle^{(0)}
  =\delta_{nm}\,\frac{C^{(0)}_{2} (\l) }{|2\sin\frac{\tau_{12}}{2} |^{2\D^{(0)}}}\ ,\qquad 
   C^{(0)}_{2}  =  \frac{\l}{8 \pi^2}  + \mc O(\l^2), \qquad
\Dz  =  1  -  \frac{\l}{8\pi^{2}}+\mc O(\l^2) \  .\la{32}
 \ee
 Expanding $\vev{\WZ}$ around $\z=1$ to  second  order gives \cite{Beccaria:2017rbe}
\ba
\la{3.3}
{\langle W^{(\z)} \rangle\ov   \vev{W^{(1)}}}\Big|_{ (\z-1)^2 } &
= (\zeta-1)^{2} \,C_{2}^{(1)}(\l)\,\frac{\pi^{3/2}\,\Gamma(-\frac{1}{2}-\gamma(\l))}{2^{1+2\gamma(\l)}\,\Gamma(-\gamma(\l))}\ , \qquad \gamma(\l) \equiv \Do  -1 \ . 
\ea
Using (\ref{1.21}),\rf{1.22}  and $\Dze =\beta'_{\z}$ in terms of  (\ref{1.7}) gives the following expressions\foot{Here we replaced the coefficients $b_{1}$, 
$b_{2}$, $b_{3}$, and $c_{1}$ by their known explicit values and used  that 
$b_{4}+b_{5}+b_{6}$ is related to $d_3$ in \rf{1.14} by \rf{1.17}.}
\ba
C_{2}^{(1)} &= \frac{\lambda}{8\pi^2}+\Big[1+8\pi^{2}(c_{2}+c_{3})\Big]\,\Big(\frac{\lambda}{4\pi^2}\Big)^{2}
+\Big[-2-\frac{\pi^{2}}{12}  + \frac{7\pi^{4}}{720}    +24\pi^{2}(c_{2}+c_{3})
\Big]\,\Big(\frac{\lambda}{4\pi^2}\Big)^{3}+\mc O(\l^{4}).
\ea
From the analogue of (\ref{3.3}) around $\z=0$, we similarly obtain 
\ba
C_{2}^{(0)} &= \frac{\lambda}{8\pi^2}+\Big(-\frac{1}{2}+8\pi^{2}c_{2}\Big)\Big(\frac{\lambda}{4\pi^2}\Big)^{2}
+\Big(\frac{1}{4}-\frac{\pi^{2}}{48}-4\pi^{2}c_{2}+64\pi^{6}b_{4}\Big)\,\Big(\frac{\lambda}{4\pi^2}\Big)^{3}+\mc O(\l^{4}).
\ea
While $\Dz $ and $\Do$ are scheme independent, the 
 $\l^{2}$ and higher terms in $C_{2}$ at $\z=0,1$ thus have scheme-dependent coefficients (cf. \rf{h1}). 
 
A similar analysis can be done for  the three-point functions. 
The three-point function of $\phi=\phi_{6}$ at $\z=1$   has the general  form   (\cf also 
\ci{Kim:2017sju,Kim:2017phs}) 
\be
\llangle\phi(\tau_1)\,\phi(\tau_2)\, \phi(\tau_3) \rrangle^{(1)}
=  { C^{(1)}_3 (\l) \ov  |2\sin\frac{\tau_{12}}{2} |^{\D(1)} \  |2\sin\frac{\tau_{23}}{2} |^{\D(1)}\  |2\sin\frac{\tau_{31}}{2} |^{\D(1)}
}\ , 
\ee
where $\Do$ is the same as in (\ref{3.1}),\rf{113}. The three-point function of 
any  particular  scalar $\phi_n $ at $\z=0$ (i.e. not coupled to the WL) is  zero due to the $\phi\to -\phi$ symmetry
of the SYM action. 
Expanding the ratio $\vev{\WZ}/\vev{W^{(1)}}$ around $\z=1$ as in \rf{3.3} 
 to  third  order gives 
\ba
{\langle W^{(\z)}\rangle\ov  \langle W^{(1)}\rangle}\Big|_{ (\z-1)^3 }  =  (\z-1)^3  \, C^{(1)}_3(\l) \, 
\frac{\pi^{3/2}\, [\Gamma(-\frac{\g}{2})]^3 \   \Gamma(-\frac{1}{2}- {3\ov 2} \gamma)}{3\cdot  2^{1+3\gamma}\, [\Gamma(-\gamma)]^3}  \ . 
\ea
Using again (\ref{1.21}) and  $\Dze =\beta'_{\z}$  from   (\ref{1.7})  gives 
\ba
C_{3}^{(1)} & = -\frac{3}{16}\Big(\frac{\lambda}{4\pi^2}\Big)^{2}+\Big[-\frac{1}{4}-\pi^{2}(3c_{2}+7c_{3})\Big]\Big(\frac{\lambda}{4\pi^2}\Big)^{3}\no \\
&\qquad\qquad  +\Big[\frac{3}{2}+\frac{9\pi^{2}}{128}-3\pi^{2}(3c_{2}+7c_{3})\Big]\Big(\frac{\lambda}{4\pi^2}\Big)^{4}+\mc O(\l^{5}),
\ea
where  the $\l^{3}$ and higher coefficients are thus  scheme dependent.

\fi


\subsection{Two-point function for ``transverse'' scalar}

Using the scalar  ladder approximation let us compute the defect two-point function for one of the  ``transverse''  scalars  $\phi_{1},...,\p_5$  (to be denoted  by   $\wt \phi$) 
which is not coupled to the scalar  Wilson line operator   defined  in \rf{2.3} 
\be
\la{pp2}
\wt G(\tau) = \frac{\vev{\tr\big[\text{P}\, \wt\phi(0)\,\wt\phi(\tau)\,\exp\int_{-L}^{L}d\tau'\, \phi(\tau')\big]}}{\vev{\tr\big[\text{P}\,\exp\int_{-L}^{L}d\tau'\, \phi(\tau')\big]}} \ . 
\ee
Here  we  shall consider an infinite straight line with $L$  as an IR cutoff.
We shall treat $\wt \phi$ and $\phi$  on equal footing  rescaling  both by $\z$, i.e.
the  averaging is done   with the  free scalar action  \rf{act} where   now 
$S = \frac{N}{\xi}\,\int d^4 x\  \text{Tr} (\partial \phi \partial \phi +   \partial\wt \phi \partial \wt\phi   )$. Then $\xi$ will  appear in the propagators of both $\phi$ and $\wt \phi$  as in \rf{2.4}.
 Alternative equivalent  option is  to  rescale  both fields in \rf{pp2}   by $\sqrt \xi$ 
 getting   the  factor  $\sqrt \xi$ in the exponents in \rf{pp2},  $\xi$-independent propagator
 and the  overall factor of $\xi$ in \rf{pp2}.  It  is then natural to remove it by 
 rescaling \be
 \la{res}
 \wt G \ \ \to \ \    \xi^{-1} \wt G\ . \ee
 The renormalization of 
 $\wt G$ and $\xi^{-1}  \wt G$ will differ  just  by a $Z$-factor  corresponding to $\xi$, i.e.
 they will satisfy closely related  Callan-Symanzik equations  (see below). 

 
At the tree and one loop level  we  will then  have  from \rf{pp2} 
\ba
\la{3.11}
& \vev{\tr\big[\text{P}\,\exp\int_{-L}^{L}d\tau'\, \phi(\tau')\big]} = 1+
\begin{tikzpicture}[line width=1 pt, scale=0.5, baseline=0]
\node[left] at (0,0) {$-L$};
\node[right] at (6,0) {$L$};
\draw[line width=2 pt] (0,0)--(6,0);
\draw (1,0) arc(180:0:2);
\node[below] at (1,0) {$\tau_{1}$};
\node[below] at (5,0) {$\tau_{2}$};
\end{tikzpicture}+\cdots\ ,\\
\la{3.10}
& \vev{\tr\big[\text{P}\, \wt\phi(0)\,\wt\phi(\tau)\,\exp\int_{-L}^{L}d\tau'\, \phi(\tau')\big]} = 
\begin{tikzpicture}[line width=1 pt, scale=0.5, baseline=0]
\node[left] at (0,0) {$-L$};
\node[right] at (6,0) {$L$};
\draw[line width=2 pt] (0,0)--(6,0);
\draw[densely dashed] (1,0) arc(180:0:2);
\node[below] at (1,0) {$0$};
\node[below] at (5,0) {$\tau$};
\end{tikzpicture}\lp
+\begin{tikzpicture}[line width=1 pt, scale=0.5, baseline=0]
\node[left] at (0,0) {$-L$};
\node[right] at (7,0) {$L$};
\draw[line width=2 pt] (0,0)--(7,0);
\draw[densely dashed] (1,0) arc(180:0:1);
\node[below] at (1,0) {$0$};
\node[below] at (3,0) {$\tau$};
\draw (4,0) arc(180:0:1);
\node[below] at (4,0) {$\tau_{1}$};
\node[below] at (6,0) {$\tau_{2}$};
\end{tikzpicture}
+\begin{tikzpicture}[line width=1 pt, scale=0.5, baseline=0]
\node[left] at (0,0) {$-L$};
\node[right] at (7,0) {$L$};
\draw[line width=2 pt] (0,0)--(7,0);
\draw (1,0) arc(180:0:1);
\node[below] at (1,0) {$\tau_{1}$};
\node[below] at (3,0) {$\tau_{2}$};
\draw[densely dashed] (4,0) arc(180:0:1);
\node[below] at (4,0) {$0$};
\node[below] at (6,0) {$\tau$};
\end{tikzpicture} \lp
+\begin{tikzpicture}[line width=1 pt, scale=0.5, baseline=0]
\node[left] at (0,0) {$-L$};
\node[right] at (6,0) {$L$};
\draw[line width=2 pt] (0,0)--(6,0);
\draw[densely dashed] (1,0) arc(180:0:2);
\node[below] at (1,0) {$0$};
\node[below] at (5,0) {$\tau$};
\draw (2,0) arc(180:0:1);
\node[below] at (2,0) {$\tau_{1}$};
\node[below] at (4,0) {$\tau_{2}$};
\end{tikzpicture}+
\begin{tikzpicture}[line width=1 pt, scale=0.5, baseline=0]
\node[left] at (0,0) {$-L$};
\node[right] at (6,0) {$L$};
\draw[line width=2 pt] (0,0)--(6,0);
\draw (1,0) arc(180:0:2);
\node[below] at (1,0) {$\tau_{1}$};
\node[below] at (5,0) {$\tau_{2}$};
\draw[densely dashed] (2,0) arc(180:0:1);
\node[below] at (2,0) {$0$};
\node[below] at (4,0) {$\tau$};
\end{tikzpicture}+\cdots\ . 
\ea
Using the propagator \rf{2.14} and  explicit expressions in (\ref{pp1})--(\ref{u1}), we get
\ba
\wt G(\tau) &= \frac{\tau^{-2+\eps } \xi }{8 \pi ^2}+\frac{\tau^{-2+\eps } ((L-\tau)^{\eps 
}+2 \tau^{\eps }-(L+\tau)^{\eps }) \xi ^2}{64 \pi ^4 (-1+\eps ) \eps 
}+\mc O(\xi ^3).
\ea
Here $\xi$ is the bare coupling. 
Applying the renormalization, i.e. the redefinition of $\xi$  as in  (\ref{2.24})-(\ref{2.25}), we find a finite result
\be\la{3311}
\wt G^{\rm ren}(\tau;\mu) = \frac{\xi}{8\pi^{2}}\frac{1}{\tau^{2}}\,\Big[1-\frac{\xi}{4\pi^{2}}\Big(1+\frac{1}{2}\log\frac{L-\tau}{L+\tau}+\log(\mu\tau)\Big)+\mc O(\xi^{2})\Big],
\ee
where $\xi$ is now the renormalized coupling $\xi(\mu)$
and the limit $L\to\infty$  is straightforward.
\iffa 
Taking $L\to\infty$ this can be put into the form 
\be
\wt G^{\rm ren}(\tau;\mu) = \mu^{2}\,\frac{C(\xi)}{8\pi^{2}}\frac{1}{(\mu\tau)^{2\wt \Delta}}, \qquad C(\xi) = \xi-\frac{\xi^{2}}{4\pi^{2}}+\cdots, \qquad \wt \Delta = 1+\frac{\xi}{8\pi^{2}}+\cdots.
\ee
\fi 
Similarly, at two loops, from  the results in (\ref{u2})--(\ref{u3}) and using again the redefinition (\ref{2.24})-(\ref{2.25}), we find a finite expression   which in the  limit  $L\to \infty$ gives 
\be
\la{3.38}
\wt G^{\rm ren}(\tau; \mu) = \frac{\xi}{8\pi^{2}}\frac{1}{\tau^{2}}\,\Big[1-\frac{\xi}{4\pi^{2}}\Big(1+\log(\mu\tau)\Big)+
\Big(\frac{\xi}{4\pi^{2}}\Big)^{2}\Big(2+\frac{\pi^{2}}{24}+\frac{5}{2}\log(\mu \tau)+\log^{2}(\mu\tau) \Big)+
\mc O(\xi^{3})\Big].
\ee
 Note that   since $\wt G^{\rm ren}(\tau)$   requires only  the renormalization of $\xi$  
 for its finiteness (i.e. no extra $Z$-factor) it satisfies 
 \be 
\Big(\mu\frac{\partial}{\partial\mu}+\beta_{\xi}\frac{\partial}{\partial\xi}\Big)\wt G^{\rm ren}(\tau; \mu) = 0\ .
 \la{cas}
 \ee 
 Since   for  generic 
 $\xi$   its   beta-function is  non-zero, i.e. the conformal  invariance is broken 
   in the ladder theory in  $d=4$,  this correlator cannot be put into the 
   standard conformal  form. 
   
A way to  achieve  conformal invariance is to  consider the  scalar ladder 
theory in  $d=4-\eps$ 
 and  specify to 
  the Wilson-Fisher fixed point \rf{2.34}. 
Considering  one loop order we have to 
keep   the full $\eps$ dependence of the tree and one-loop terms.  
Then  instead of (\ref{3.38}) we get
\ba
\wt G^{\rm ren}(\tau) = &\frac{\mu^{\eps}}{2\,\tau^{2-\eps}}\, \frac{\xi}{4\pi^{2}}+
\frac{\mu ^{\eps } (2-2 \eps -\mu ^{\eps } 
(L-\tau )^{\eps }-2 \mu ^{\eps } \tau ^{\eps }+\mu ^{\eps } (L+\tau 
)^{\eps })}{4 \tau ^{2-\eps }  (1-\eps ) \eps }  \Big(\frac{\xi}{4\pi^{2}}\Big)^{2} \lp
+\frac{1}{\tau^{2}}\,\left(1+\frac{\pi^{2}}{48}+\frac{5}{4}\log(\mu \tau)+\frac{1}{2}\log^{2}(\mu\tau)\right)\,\Big(\frac{\xi}{4\pi^{2}}\Big)^{3}
+\cdots\ .
\ea
Evaluating this at $\xi=\xi^{*}$ in \rf{2.34}, expanding in $\eps$, and taking the infinite line limit $L\to\infty$ gives simply
\be
\wt G^{\rm ren}(\tau)\Big|_{\xi=\xi^{*}} = \frac{1}{2\tau^{2}}\,\Big(\eps-\frac{1}{2}\,\eps+\frac{\pi^{2}}{24}\,\eps^{3}+\cdots\Big).
\ee
The fact that all $\log \tau$ terms  cancel implies that the scaling with $\tau$ is
 non-anomalous at the conformal fixed point 
 (i.e.  the corresponding dimension is $\wt \Delta^*=1$), at least to  two loops. 

This  is analogous  with what happens in full  $\N=4$  SYM  theory case, where the insertion  of the  ``transverse'' scalars
into the WML  is BPS-protected.
  Here in the ladder model   we do not  have an argument
based on supersymmetry,  but we  can explain  why  $\wt \Delta^{*}=1$ as follows. If one starts with the WL  coupled to two scalars as 
\be\la{319}
\tr\text{P}\exp\int d\tau \, (\z_{1}\phi+\z_{2}\wt \phi) \ , 
\ee
then  setting  $\z_{1}=\z \cos\alpha$, $\z_{2}=\z \sin\alpha$   and 
  redefining  the scalars
 (using $SO(6)$  symmetry of the  bulk  scalar action) 
 we get back to the  WL coupled to a single scalar with the coupling $\z$.
  Thus the  beta-function 
cannot be a function of $\alpha$, but only of $\z$, i.e. 
the angle $\alpha$ should be a parameter of an  exactly marginal deformation.
For infinitesimal $\alpha$  the  integrand in the exponent in \rf{319} 
is $\z \phi + \z \alpha\, \wt \phi$
so that the insertion of $\wt \phi$ into the WL coupled to $\phi$ should be exactly marginal with $\Delta=1$ at a  fixed point.

\def \ve  {\varepsilon}

\subsection{Two-point function for   coupled  scalar}

The same calculation for the scalar  coupled to the loop, \ie for 
\be\la{321}
G(\tau) = \frac{\vev{\tr\big[\text{P}\, \phi(0)\, \phi(\tau)\,\exp\int_{-L}^{L}d\tau'\, \phi(\tau')\big]}}{\vev{\tr\big[\text{P}\,\exp\int_{-L}^{L}d\tau'\, \phi(\tau')\big]}},
\ee
requires us  to consider all possible contractions, including those involving the fields at 0 and $\tau$.
At tree and one-loop orders we then get the following contributions 
\ba
& \vev{\tr\big[\text{P}\, \phi(0)\, \phi(\tau)\,\exp\int_{-L}^{L}d\tau'\, \phi(\tau')\big]} = 
\begin{tikzpicture}[line width=1 pt, scale=0.5, baseline=0]
\draw[line width=2 pt] (0,0)--(6,0);
\draw (1.5,0) arc(180:0:1.5);
\node[below] at (1.5,0) {$0$};
\node[below] at (4.5,0) {$\tau$};
\end{tikzpicture}\lp\qquad \qquad 
+\int d^2 \tau \,\Big[
\begin{tikzpicture}[line width=1 pt, scale=0.5, baseline=0]
\draw[line width=2 pt] (0,0)--(7,0);
\draw (1,0) arc(180:0:1);
\draw (4,0) arc(180:0:1);
\end{tikzpicture}
+\begin{tikzpicture}[line width=1 pt, scale=0.5, baseline=0]
\draw[line width=2 pt] (0,0)--(6,0);
\draw (1,0) arc(180:0:2);
\draw (2,0) arc(180:0:1);
\end{tikzpicture}\Big]+\cdots\ .
\ea
Each structure splits into 6 diagrams depending on the order of $\tau'<\tau''$ with respect to $0<\tau$. The result is 
\ba
\int d^2 \tau \, & 
\begin{tikzpicture}[line width=1 pt, scale=0.5, baseline=0]
\draw[line width=2 pt] (0,0)--(7,0);
\draw (1,0) arc(180:0:1);
\draw (4,0) arc(180:0:1);
\end{tikzpicture} = \frac{\xi ^2 (L (L-\tau ))^{-1+\eps }}{64 \pi ^4 (-1+\eps 
)^2}+\frac{L^{\eps } \xi ^2 \tau ^{-2+\eps }}{64 \pi ^4 (-1+\eps ) 
\eps }+\frac{\xi ^2 (L-\tau )^{\eps } \tau ^{-2+\eps }}{64 \pi ^4 
(-1+\eps ) \eps }\lp
+\frac{\xi ^2 (L \tau )^{-1+\eps }}{64 \pi ^4 
(-1+\eps )^2}+\frac{\xi ^2 ((L-\tau ) \tau )^{-1+\eps }}{64 \pi ^4 
(-1+\eps )^2}+\frac{\xi ^2 \tau ^{-2+2 \eps } \Gamma (-1+\eps ) 
\Gamma (\eps )}{64 \pi ^4 (-1+\eps ) \Gamma (-1+2 \eps )}, \\
\int d^2 \tau\, & \begin{tikzpicture}[line width=1 pt, scale=0.5, baseline=0]
\draw[line width=2 pt] (0,0)--(6,0);
\draw (1,0) arc(180:0:2);
\draw (2,0) arc(180:0:1);
\end{tikzpicture} = \frac{\xi ^2 \tau ^{-2+2 \eps }}{64 \pi ^4 (-1+\eps ) \eps 
}
 +\frac{\xi ^2 \tau ^{-2+\eps } (L^{\eps } \tau -L \tau ^{\eps })}{64 
L \pi ^4 (-1+\eps )^2}\lp 
  +\frac{\xi ^2 \tau ^{-2+\eps } ((-1+2^{\eps }) 
L^{\eps }
+\tau ^{\eps }-(L+\tau )^{\eps })}{64 \pi ^4 (-1+\eps ) \eps 
}
+\frac{\xi ^2 \tau ^{-2+\eps } (-\tau ^{\eps } (L+\tau )+\tau  (L+\tau )^{\eps })}{64 \pi ^4 (-1+\eps )^2 (L+\tau )} \lp  -\frac{\xi ^2 \tau \
^{-2+2 \eps } B_{\frac{\tau }{L}}(2-2 \eps ,\eps )}{64 \pi ^4 
(-1+\eps )}
-\frac{2^{-5-2 \eps } \xi ^2 \tau ^{-2+2 \eps } \Gamma 
(\frac{3}{2}-\eps ) \Gamma (-1+\eps )}{\pi ^{9/2} (-1+\eps 
)}\lp +\frac{\xi ^2 ((L (L+\tau ))^{-1+\eps }-(L \tau )^{-1+\eps } \, 
_2F_1(1-\eps ,-1+\eps ;\eps ;-\frac{L}{\tau }))}{64 \pi ^4 (-1+\eps 
)^2}.
\ea
Taking  $L\to\infty$ limit  in the $B$-function,\footnote{Here $B$ is  the incomplete 
beta function $B_{z}(\alpha, \beta) = \int_{0}^{z}dw\, w^{\alpha-1}(1-w)^{\beta-1}$   
with $B_{0}(\alpha,\beta)=0$.}    using  
\be
_2F_1(1-\eps, -1+\eps, \eps; -x) = x^{1-\eps}\frac{\Gamma(2-2\eps)\Gamma(\eps)}{\Gamma(1-\eps)}+\mc O(x^{-1+\eps}), \qquad x\to +\infty, 
\ee
 and   redefining  $\xi$  according to \rf{2.24} we find that the 
  renormalization of $G(\tau) $ in \rf{321} requires also an additional $Z$-factor 
\be
\la{3.50}
G^{\rm ren}(\tau; \mu) = \lim_{L\to\infty }\lim_{\eps\to 0}Z\,G(\tau), \qquad\qquad  Z = 1+\frac{1}{2\pi^{2}}\frac{\mu^{-\eps}\,\xi}{\eps}+\cdots\ . 
\ee
As a result, we find
\be
G^{\rm ren}(\tau;\mu) = \mu^{2}\frac{\xi}{8\pi^{2}\,(\mu \tau)^{2-\eps}}\Big[1+(-2-3\log(\mu\tau))\frac{\xi}{4\pi^{2}}+\cdots\Big],
\ee
where we restored  the  full  $\eps$-dependence of  the tree-level term.
Replacing now  $\xi$   by its  Wilson-Fischer  fixed point   value 
 $\xi^{*}=4\pi^{2}\eps+2\pi^{2}\eps^{2}+\cdots$ in \rf{2.34} 
 we find
\be
G^{*}(\tau;\mu) = \frac{\eps}{2\tau^{2}}\big[1+(-2-2\log(\mu \tau))\big]\,\eps+\mc O(\eps^{2}).
\ee
This has the conformal form with a non-trivial dimension 
 $\Delta^{*} = 1+\eps+...$, in agreement with the expected relation 
between the anomalous dimension and the derivative of the beta-function in \rf{2.33}
\be
\Delta^{*} = 1+\beta'_{\xi^{*}} = 1+\eps-\frac{1}{2}\eps^{2}+\cdots.
\ee
At the two-loop order  we need    the following  contributions
\ba
\int d^{4}\tau\Big[ & 
\begin{tikzpicture}[line width=1 pt, scale=0.5, baseline=0]
\draw[line width=2 pt] (0,0)--(10,0);
\draw (1,0) arc(180:0:1);
\draw (4,0) arc(180:0:1);
\draw (7,0) arc(180:0:1);
\end{tikzpicture}+
\begin{tikzpicture}[line width=1 pt, scale=0.5, baseline=0]
\draw[line width=2 pt] (0,0)--(7,0);
\draw (1,0) arc(180:0:1);
\draw(4,0) arc(180:0:1);
\draw(4.5,0) arc(180:0:0.5);
\end{tikzpicture}
+\begin{tikzpicture}[line width=1 pt, scale=0.5, baseline=0]
\draw[line width=2 pt] (0,0)--(7,0);
\draw (1.5,0) arc(180:0:0.5);
\draw(1,0) arc(180:0:1);
\draw (4,0) arc(180:0:1);
\end{tikzpicture}\nonumber \\
& + \begin{tikzpicture}[line width=1 pt, scale=0.5, baseline=0]
\draw[line width=2 pt] (0,0)--(6,0);
\draw  (1,0) arc(180:0:2);
\draw(1.5,0) arc(180:0:1.5);
\draw(2,0) arc(180:0:1);
\end{tikzpicture}
+ \begin{tikzpicture}[line width=1 pt, scale=0.5, baseline=0]
\draw[line width=2 pt] (0,0)--(6,0);
\draw  (1,0) arc(180:0:2);
\draw(1.25,0) arc(180:0:0.75);
\draw(3.25,0) arc(180:0:0.75);
\end{tikzpicture}
\Big].
\ea
Each diagram correspond to 15 possible orderings of the positions on the line, two fixed at $0$ and $\tau$,  and four  at  
 various  possible positions 
consistent  with the planarity constraint.
To 
extract  divergences from closed expressions  for  the diagrams   here   we use 
  a simple   cutoff  regularization
\be
D(\tau)\to D_{a}(\tau) = \frac{\xi}{8\pi^{2}}\frac{1}{(|\tau|+a)^{2}} \ , \qquad \qquad a\to 0 \ . 
\ee
Then one finds, for example, that 
\ba
\int d^{4}\tau \begin{tikzpicture}[line width=1 pt, scale=0.5, baseline=0]
\draw[line width=2 pt] (0,0)--(6,0);
\draw  (1,0) arc(180:0:2);
\draw(1.25,0) arc(180:0:0.75);
\draw(3.25,0) arc(180:0:0.75);
\end{tikzpicture} = &L\,\frac{(-2 a+\tau) \xi ^3 \log 2}{256 a \pi ^6 \tau^3}
-\frac{\xi ^3 (-1+\log 256+16 \log \frac{\tau}{L})}{1024 \pi ^6 
a^2}\lp
+\frac{\xi ^3 (-8+\log 32+20 \log \frac{a}{L}-12 \log 
\frac{\tau}{L})}{512 \pi ^6 \tau a}\\ &
+\frac{\xi ^3}{1024 \
\pi ^6 \tau^2} \Big[-19+\pi ^2-12 \log 2-16 \log \frac{a}{L}-8 \log 2 \log \frac{a}{L}\lp
+12 \log ^2 \frac{a}{L} -4 \log \frac{\tau}{L}+2 \log 4 \log \frac{\tau}{L}-32 
\log \frac{a}{L} \log \frac{\tau}{L}+18 \log ^2\frac{\tau}{L}\Big]+\cdots\nonumber
\ea
where we already  expanded in large $L$ and small $a$ (before  the expansion the expression is unwieldy).
Recomputing the one-loop diagrams with this regularization and adding the two-loop ones, one finds that  one can define the renormalized  correlator as\footnote{The order of limits here  is important.} 
\be
G^{\rm ren}(\tau) = \lim_{L\to\infty }\lim_{a\to 0}Z\,G(\tau)\ .  
\ee
The bare coupling $\xi(a)$  appearing in the expansion of $G$ is related to the renormalized coupling $\xi=\xi(\mu)$
at the  inverse length scale $\mu$
 by 
\ba
\la{3.61}
\xi (a) = \xi -\log(a \, \mu)\,\frac{\xi^{2}}{4\pi^{2}}+\Big[\frac{1}{2}\log(a\,\mu)+\log^{2}(a\,\mu)\Big]\,\frac{\xi^{3}}{(4\pi^{2})^{2}}+\cdots, 
\ea
corresponding to the beta-function in \rf{2.27}. The renormalization factor $Z$ is  given by 
\be\la{zz}
Z = 1-2\log(a\,\mu)\,\frac{\xi^{2}}{4\pi^{2}}+[2\,\log(a\,\mu)+\log^{2}(a\,\mu)]\,\frac{\xi^{2}}{(4\pi^{2})^{2}}+\cdots\ .
\ee
The renormalized two-point function  then reads
\be
G^{\rm ren}(\tau;  \mu) = \frac{1}{2\,\tau^{2}}\frac{\xi}{4\pi^{2}}\,\Big[
1+\Big(1-3\log(\mu\tau)\Big)\frac{\xi}{4\pi^{2}}+\Big(-2+\frac{5\pi^{2}}{24}-\frac{3}{2}\log(\mu\tau)+6\log^{2}(\mu\tau)\Big)\,\frac{\xi^{2}}{(4\pi^{2})^{2}}+\cdots\Big].
\ee
Like \rf{3.38} this  expression 
 cannot be put into the  conformal form (with all $\tau$ dependence
 appearing only as a power $(\mu\tau)^{2\Delta}$)   since the conformal invariance is 
 broken for generic $\xi$. 
 
 The renormalized   correlator   satisfies 
  the Callan-Symanzik equation 
\be\la{347}
\Big[\mu\frac{\partial}{\partial\mu}+\beta_{\xi}\frac{\partial}{\partial\xi}+2(\Delta-1)\Big]\,\xi^{-1}G^{\rm ren}(\tau; \mu) = 0\ .
\ee
 We have written \rf{347} 
for the rescaled \rf{res}   correlator corresponding to the picture   where 
$\sqrt \xi$  coupling appears in the exponent of the WL operator  and the two-point function is defined   for the  canonically normalized  fields (i.e. without extra coupling factor in the bulk action compared to  \rf{act}). Thus 
 $\Delta-1$  in \rf{347}  is the anomalous dimension of the canonically   normalized  scalar $\phi$.


Using \rf{2.27}, i.e.  $\beta_{\xi}=\frac{1}{4\pi^{2}}\xi^{2}-\frac{1}{32\pi^{4}}\xi^{3}+\cdots$,  we obtain 
\be
\la{3.65}
\Delta = 1+\frac{3}{8\pi^{2}}\xi-\frac{5}{64\pi^{4}}\xi^{2}+\cdots.
\ee
This expression  is in agreement with the general relation  for the anomalous dimension 
$\Delta -1= \frac{d}{dg}\beta_{g}$ where  the coupling here  is $g=\sqrt\xi$  appearing in front of  the   scalar   $\phi$ in the  exponent, \ie
\be
\beta_{g} = \mu\frac{\partial}{\partial\mu}g = \frac{1}{2\sqrt\xi}\beta_{\xi} = \frac{1}{8\pi^{2}}g^{3}-\frac{g^{5}}{64\pi^{4}}+\cdots \ .
\ee
The  dimension (\ref{3.65})  in the effective 
 ladder model  
 corresponds to the terms with highest power of $\z$ 
 in the 
 dimension  of the scalar $\phi=\phi_6$ 
in the full  $\N=4$ SYM theory   (cf. \rf{1.8})
\be\la{340}
\Delta^{(\z)} = 1+\beta'(\z) = 1+\frac{\l}{8\pi^2}(3\z^2-1)+\frac{\l^2}{64\pi^4}\,(1-5\z^4)+\mc O(\l^{3}).
\ee
Similarly,  the  Callan-Symanzik  equation \rf{cas} for the 
two-point function  $\wt G^{\rm ren}$  for the ``transverse'' scalar $\wt \phi$  in \rf{3.38}
rewritten  for the rescaled \rf{res} correlator, i.e.  in the   form \rf{347}, is 
  is 
\be\la{3477}
\Big(\mu\frac{\partial}{\partial\mu}+\beta_{\xi}\frac{\partial}{\partial\xi}\Big)\,\wt G^{\rm ren}(\tau; \mu) =    \Big[\mu\frac{\partial}{\partial\mu}+\beta_{\xi}\frac{\partial}{\partial\xi}+2(\wt \Delta-1)\Big]\,\xi^{-1}\wt G^{\rm ren}(\tau; \mu) =      0\ . 
\ee
It implies that the anomalous 
dimension $\wt \Delta-1$ of   the canonically   normalized    scalar field $\wt \phi$
 found  in the ladder   approximation 
 is  directly proportional to the  beta-function $\beta_{\xi}$ in \rf{2.27}
\be \la{333}
\wt \Delta =1 +    {1\ov 2}\xi^{-1}  \beta_{\xi}
   = 1 + \frac{\xi}{8\pi^2}-\frac{\xi^2}{64\pi^4}+\mc O(\xi^{3}).
\ee
To reconstruct the  expression for the  corresponding  anomalous dimension $\wt\Delta-1$ in the full SYM theory 
  (i.e. the analog of \rf{340})
  which is a function of $\l$ and $\z$   we may use that\     (i)   it 
   should  reduce  to \rf{333}
if one keeps  only  highest powers of $\z$  at each order in $\l$;  (ii) it 
 should  vanish at $\z=\pm 1$; 
 (iii)  it should   be equal to \rf{340}  at $\z=0$ as then all 6 scalars are   on equal footing (not coupled to the loop). 
 This  gives 
\be\la{341}
\wt\Delta^{(\z)}=1 + \frac{\l}{8\pi^{2}}(\z^2-1)-\frac{\l^{2}}{64\pi^{4}} (\z^4-1)+\mc O(\l^{3}).
\ee
This  is  simply related  to the   beta-function $ {\beta_{\z}}$ in \rf{1.8}\foot{The relation between the prefactors in \rf{333}   and in \rf{342}   follows from 
 ${1\ov 2} \xi^{-1}  {d\ov d \mu } \xi = {1\ov 2} \z^{-2}   {d\ov d \mu }  \z^2 =\z ^{-1} \beta_\z$. 
}
\be
\la{342}
\wt\Delta^{(\z)}= 1+\zeta^{-1} \, {\beta_{\z}} \ .
\ee
Thus  while the anomalous  dimension of the coupled scalar \rf{340}  is given by the derivative of the beta-function of $\z$,  the anomalous dimension of the ``transverse'' scalar \rf{342}  is proportional to the beta-function itself. 

A way  to 
  understand   why  this  relation of $\wt \Delta^{(\z)} $ to the  beta-function 
  should hold in general  let us  start again with 
the loop \rf{319}  coupled to the two  scalars  ($\phi=\phi_6$ and a  ``transverse'' $\wt \phi$)  with different coefficients, i.e. 
\be\la{343}
\z_{1}\,\phi+\z_{2}\,\wt \phi \ , \ \ \ \ \ \qquad   \z_{1} = \z\cos\alpha, \qquad \z_{2}=\z\sin\alpha \ .
\ee
In this case we may formally define    two  beta-functions $ \beta_{\z_{i}}=\mu { d \ov d \mu} \z_i $  but since $\alpha$ is an exactly marginal parameter  not running with $\mu$  (see discussion below \rf{319})  we should  have 
\be\la{344}
 \beta_{\z_{1}} = \cos \a\, \beta_\z \ ,\qquad  \ \ \ \ \ \ \  \beta_{\z_{2}} = \sin \a\, \beta_\z \ .\ee
At the same time,   the general relations for  the  anomalous 
dimensions  of  $\p$   and $\wt \p$  at  the point $\z_2=0$ or $\a=0$  (when $\wt \p$ is not coupled to the loop) 
 are   given by 
\be\la{345} \Delta -1= \frac{\del }{\del \z_{1}} \beta_{\z_{1}}\Big|_{\z_{2}=0} =   \beta'_{\z} \ , \qquad \qquad 
\wt\Delta -1= \frac{\del}{\del\z_{2}} \beta_{\z_{2}}\Big|_{\z_{2}=0} = \z^{-1}  \beta_{\z} \ , 
\ee
where in the last equality we used that $\frac{\del}{\del\z_{2}} \beta_{\z_{2}}\big|_{\z_{2}=0} = ( \z^{-1} \cos \a { \del \ov \del \a} +   \sin \a { \del \ov \del \z})  ( \sin \a\, \b_\z )
\big|_{\alpha=0} = \z^{-1} \b_\z$.

The above   argument   implies,  in particular,  that  the fact that $\wt \Delta=1$ at the fixed points when 
 $\beta_{\z}=0$ and $\z\neq 0$  follows essentially from the
 rotational symmetry  in the scalar space.
  For the  fixed point at $\zeta=0$, 
where all scalars are not coupled to the loop   and the 
 full $SO(6)$ rotational symmetry is restored, this does not apply as here 
  all scalars have then the same dimension $\Dz=1+\beta'_{\z}|_{\z=0}$.

\section*{Acknowledgements}
We are grateful to  K. Zarembo  for very useful discussions. 
We also  thank      N. Gromov    and J. Henn   for  helpful  communications. 
MB was supported by the INFN grant GSS (Gauge Theories, Strings and Supergravity). 
The work of SG is 
supported in part by the US NSF under Grant No. PHY-1914860. AAT was 
supported by the STFC grant ST/T000791/1. 

\
\def \t {\tau}

\appendix

\section{Conventions and useful formulae \la{AA}}

For $SU(N)$ generators in the fundamental representation we have 
\be
[T^{a}, T^{b}] = i\,f^{abc}\,T^{c}, \quad \tr T^{a}=0, \quad \tr T^{a}T^{b} = \frac{1}{2}\delta^{ab},\quad (T^{a}T^{a})_{ij} =\frac{N^{2}-1}{2N}\,\delta_{ij}.
\ee
\be
T^{a}_{ij}T^{a}_{kl} = \frac{1}{2}\Big(\delta_{il}\delta_{jk}-\frac{1}{N}\delta_{ij}\delta_{kl}\Big), \qquad 
f^{acd}f^{bcd} = N\delta^{ab}.
\ee
In computing  higher-order corrections one also needs 
\ba
\tr(T^{a}T^{a}T^{b}T^{b}) &= \frac{1}{2}\tr(1)\tr(T^{b}T^{b})-\frac{1}{2N}\tr(T^{b}T^{b}) = (\frac{N}{2}-\frac{1}{2N})\frac{N^{2}-1}{2} = 
\frac{(N^{2}-1)^{2}}{4N}, \\
\tr(T^{a}T^{b}T^{a}T^{b}) &= \frac{1}{2}\tr(T^{b})\tr(T^{b})-\frac{1}{2N}\tr(T^{b}T^{b}) = -\frac{1}{2N}\frac{N^{2}-1}{2} = -\frac{N^{2}-1}{4N}, \\
\tr(T^{a}T^{b}T^{b}T^{a}) &= \tr(T^{a}T^{a}T^{b}T^{b}) = \frac{(N^{2}-1)^{2}}{4N}.
\ea
Useful formulae for diagram computations are 
\be
\int_{0}^{1}\prod_{i=1}^{N}d\alpha_{i}\, \alpha_{i}^{\nu_{i}-1}\,\delta(1-\sum_{i}\alpha_{i}) = \frac{\Gamma(\nu_{1})\cdots\Gamma(\nu_{N})}
{\Gamma(\nu_{1}+\cdots+\nu_{N})},
\ee
\be
\int d^{2\omega}w\frac{1}{(w^{2}+M^{2})^{s}} = \pi^{\omega}\frac{\Gamma(s-\omega)}{\Gamma(s)}(M^{2})^{\omega-s},
\ee
\ba
& \int_{\tau>\tau_{1}>\tau_{2}>\tau_{3}>0}  d\t_1\, d\t_2\, d\t_3\,  (\tau_{1}-\tau_{2})^{c_{12}}(\tau_{2}-\tau_{3})^{c_{23}}(\tau_{1}-\tau_{3})^{c_{13}}\lp\qquad  
\qquad = \tau^{c_{12}+c_{23}+c_{13}+3}\ \frac{\Gamma(c_{12}+1)\Gamma(c_{23}+1)}{(c_{12}+c_{23}+c_{13}+2)(c_{12}+c_{23}+c_{13}+3)\,
\Gamma(c_{12}+c_{23}+2)},
\ea
\be
\la{F.7}
\int_{0}^{1}d\alpha\int_{0}^{1-\alpha}d\beta\,(\alpha\beta)^{p}(\alpha+\beta)^{q}(1-\alpha-\beta)^{r} = \frac{2^{-1-2 p} \sqrt{\pi } \Gamma (1+p) \Gamma (2+2 p+q) \Gamma 
(1+r)}{\Gamma (\frac{3}{2}+p) \Gamma (3+2 p+q+r)}\ . 
\ee
Let us  also  recall   some  general relations  between pole coefficients   and beta-function in dimensional regularization. 
If  $\xi(\eps)$ is a bare coupling    and  $\xi(\mu)$ is a renormalized one 
we have 
\be\la{b1}
\xi (\eps) = \mu^{\eps}\Big[\xi(\mu)+\frac{T_{1}(\xi(\mu))}{\eps}+\frac{T_{2}(\xi(\mu))}{\eps^{2}}+\cdots\Big],
\ee
where $T_{n}(\xi)$ have   perturbative expansions 
\ba
T_{1}(\xi) &= p_{11}\xi^{2}+p_{21}\xi^{3}+p_{31}\xi^{4}+p_{41}\xi^{5}+\cdots, \qquad
T_{2}(\xi) = p_{22}\xi^{3}+p_{32}\xi^{4}+p_{42}\xi^{5}+\cdots, \no\\ 
T_{3}(\xi) &= p_{33}\xi^{4}+p_{43}\xi^{5}+\cdots, \qquad \qquad \qquad \qquad \ \ \ 
T_{4}(\xi) = p_{44}\xi^{5}+\cdots,\ \ \   .... \ .\la{b3}
\ea
Differentiating  \rf{b1} over $\mu$,   using   $\mu\frac{d}{d\mu}\xi(\eps)=0$ and   setting to zero  the resulting coefficients of poles in $\eps$  
gives differential constraints  on $T_n$ 
\ba
& -T_{2}+\xi T_{2}'+T_{1}T_{1}'-\xi T_{1}'^{2} = 0, \\
& -T_3+T_2 T_1'-T_1 T_1'{}^2+\xi  
T_1'{}^3+T_1 T_2'-2 \xi  T_1' T_2'+\xi  
T_3'=0, \\
& -T_4+T_3 T_1'-T_2 T_1'{}^2+T_1 
T_1'{}^3-\xi  T_1'{}^4\lp \qquad +T_2 T_2'-2 T_1 
T_1' T_2'+3 \xi  T_1'{}^2 T_2'-\xi  T_2'{}^2 
+T_1 T_3'-2 \xi  T_1' T_3'+\xi  T_4' = 0, \ \ \  \  ....  \ . 
\ea
Plugging here  the expansions \rf{b3}
gives coefficients of all higher poles in terms of the  coefficients  of the simple one
\ba
\la{B.9}
\te  p_{22} = p_{11}^{2}, \ \ 
p_{32} = \frac{7}{3}\,p_{11}p_{21}, \ \ 
p_{33} = p_{11}^{3}, \quad 
p_{42} = \frac{3}{2}p_{21}^{2}+\frac{5}{2}p_{11}p_{31}, \ \ 
p_{43} = \frac{23}{6} p_{11}^{2}p_{21}, \ \ 
p_{44} = p_{11}^{4}, \   ....
\ea
The beta-function can be expressed in terms of $T_{1}$  as  
\be
\la{B.10}
\beta(\xi)= \mu { d \ov d \mu } \xi(\mu)  = -T_{1}+\xi T_{1}' = p_{11}\xi^{2}+2\,p_{21}\xi^{3}+3\,p_{31}\xi^{4}+4\,p_{41}\xi^{5} +  \cdots.
\ee

\section{Computation of Wilson loop on a circle in  ladder approximation}
\la{A}

Let us illustrate how to use the loop equation \rf{2.5}  to automatically generate relevant   ladder diagrams in the planar  expansion on the  example 
of  the two-loop calculation of the WL in \rf{1.21}. 
Following the mode regularization approach in \cite{Beccaria:2018ocq}, 
we replace the  propagator on the circle  in (\ref{2.4}) by  its regularized version 
\be
\la{A.1} 
D_\eps(\tau) = \frac{\xi}{8\,\pi^2}\,\sum_{n=1}^\infty e^{-n\,\ve}\,(-n)\,\cos(n\,\t)\  ,
\ee
where $\ve = { a \ov R} \to 0$  and $a$  is a  cutoff of dimension of length. 
Using this in the loop equation \rf{2.5}, the first-order term in the expansion (\ref{2.15}) is simply
\be
\la{A.2}
\GW _1(\t) = \sum_{n=1}^\infty e^{-n\,\ve}\frac{\cos(n \t)-1}{8\,\pi^2\,n}.
\ee
Setting  $\t\to 2\pi$ before summing over $n$, one gets 
\be
\GW _1(2\pi)=0.
\ee
Next, using (\ref{A.2})  in the loop equation, we obtain 
\ba
\la{A.4}
\GW _2(\t) = &\sum_{n_1,n_2=1}^\infty e^{-\ve (n_1+n_2)}  \Big\{
\frac{n_2^2  \sin (n_1 \t) \sin (n_2
   \t)}{32 \pi ^4 (n_1^2-n_2^2){}^2}  +\cos (n_1 \t)
   \Big[\frac{(n_1^2+n_2^2) n_2  \cos (n_2
   \t)}{64 \pi ^4 n_1 (n_1^2-n_2^2){}^2} \no \\
& \qquad \qquad  +\frac{n_2 }{64 \pi ^4 (n_1^3-n_1 n_2^2)}\Big] 
  +\frac{(n_2^2-2 n_1^2)
    \cos (n_2 \t)}{64 \pi ^4 n_1
   (n_1^2-n_2^2) n_2}+\frac{(2 n_1^4-5 n_2^2 n_1^2+n_2^4)
   }{64 \pi ^4 n_1 (n_1^2-n_2^2){}^2 n_2} \Big\} \ . 
\ea
Here  
  the special case
$n_1=n_2$ should be treated separately and   one gets 
\begin{align}
\GW _2(\t) &= \sum_{n_1\neq n_2=1}^\infty(\cdots)+\sum_{n=1}^\infty
e^{-2 n \ve  } \Big[\frac{15-2 n^2 \t^2}{512 \pi ^4 n^2}-\frac{\cos (n \t)}{32 \pi ^4
   n^2}+\frac{\cos (2 n \t)}{512 \pi ^4 n^2}-\frac{\t \sin (n \t)}{128 \pi ^4 n}\Big],
\end{align}
where the first term in the r.h.s. vanishes for $\t=2\pi$.  As a  result,   
\be
\GW _2(2\pi) = -\frac{1}{64\pi^2}\sum_{n=1}^\infty e^{-2n\ve } = -\frac{1}{128\pi^2\ve }+
\frac{1}{128\pi^2}+\mc O(\ve ).
\ee
Dropping the  singular term (linear divergence),  we  reproduce  the ladder part of  the  expression  in \rf{1.21},\rf{1.26} (cf. \rf{21}) 
\be
{W} = 1+0\,\cdot \xi+\frac{1}{128\pi^2}\,\xi^{2}+\dots \ . 
\ee

\section{Contributions to  two-point  function for ``transverse'' scalar  
 to  two loops}
\la{app:diagrams}

Using the propagator in (\ref{2.14}), we have the following explicit expressions for the one loop diagrams in (\ref{3.10}) and (\ref{3.11})
\ba
\la{pp1}
\begin{tikzpicture}[line width=1 pt, scale=0.5, baseline=0]
\node[left] at (0,0) {$-L$};
\node[right] at (6,0) {$L$};
\draw[line width=2 pt] (0,0)--(6,0);
\draw (1,0) arc(180:0:2);
\node[below] at (1,0) {$0$};
\node[below] at (5,0) {$\tau$};
\end{tikzpicture} &= \frac{\xi}{8\pi^{2}}\frac{1}{\tau^{2-\eps}}, \\
\begin{tikzpicture}[line width=1 pt, scale=0.5, baseline=0]
\node[left] at (0,0) {$-L$};
\node[right] at (7,0) {$L$};
\draw[line width=2 pt] (0,0)--(7,0);
\draw (1,0) arc(180:0:1);
\node[below] at (1,0) {$0$};
\node[below] at (3,0) {$\tau$};
\draw (4,0) arc(180:0:1);
\node[below] at (4,0) {$\tau_{1}$};
\node[below] at (6,0) {$\tau_{2}$};
\end{tikzpicture} &= \frac{(L-\tau)^{\eps } \tau^{-2+\eps }}{64 \pi ^4 (-1+\eps ) \eps } \xi ^2,  \\
\begin{tikzpicture}[line width=1 pt, scale=0.5, baseline=0]
\node[left] at (0,0) {$-L$};
\node[right] at (7,0) {$L$};
\draw[line width=2 pt] (0,0)--(7,0);
\draw (1,0) arc(180:0:1);
\node[below] at (1,0) {$\tau_{1}$};
\node[below] at (3,0) {$\tau_{2}$};
\draw (4,0) arc(180:0:1);
\node[below] at (4,0) {$\tau_{0}$};
\node[below] at (6,0) {$\tau$};
\end{tikzpicture} &=\frac{L^{\eps } \tau^{-2+\eps } }{64 \pi ^4 (-1+\eps ) \eps }\,\xi ^2,  \\
\begin{tikzpicture}[line width=1 pt, scale=0.5, baseline=0]
\node[left] at (0,0) {$-L$};
\node[right] at (6,0) {$L$};
\draw[line width=2 pt] (0,0)--(6,0);
\draw (1,0) arc(180:0:2);
\node[below] at (1,0) {$0$};
\node[below] at (5,0) {$\tau$};
\draw (2,0) arc(180:0:1);
\node[below] at (2,0) {$\tau_{1}$};
\node[below] at (4,0) {$\tau_{2}$};
\end{tikzpicture} &= \frac{\tau^{-2+2 \eps } }{64 \pi ^4 (-1+\eps ) \eps }\,\xi ^2,  \\
\begin{tikzpicture}[line width=1 pt, scale=0.5, baseline=0]
\node[left] at (0,0) {$-L$};
\node[right] at (6,0) {$L$};
\draw[line width=2 pt] (0,0)--(6,0);
\draw (1,0) arc(180:0:2);
\node[below] at (1,0) {$\tau_{1}$};
\node[below] at (5,0) {$\tau_{2}$};
\draw (2,0) arc(180:0:1);
\node[below] at (2,0) {$0$};
\node[below] at (4,0) {$\tau$};
\end{tikzpicture} &= \frac{\tau^{-2+\eps } ((-1+2^{\eps }) L^{\eps }+\tau^{\eps }-(L+\tau)^{\eps }) 
}{64 \pi ^4 (-1+\eps ) \eps }\,\xi ^2,  \\
\begin{tikzpicture}[line width=1 pt, scale=0.5, baseline=0]
\node[left] at (0,0) {$-L$};
\node[right] at (6,0) {$L$};
\draw[line width=2 pt] (0,0)--(6,0);
\draw (1,0) arc(180:0:2);
\node[below] at (1,0) {$\tau_{1}$};
\node[below] at (5,0) {$\tau_{2}$};
\end{tikzpicture} &= \frac{2^{-3+\eps } L^{\eps } }{\pi ^2 (-1+\eps ) \eps }\, \xi. \la{u1}
\ea
At two loops, 
we  have 15 diagrams $N_{i}$, $i=1, \dots, 15$, in the numerator of (\ref{pp2})
and two diagrams $D_{1}$, $D_{2}$ in the denominator. Their expressions are\foot{Here we omit the  labels $\pm L$  at  the  ends of the line.}
\be
\la{u2}
N_{1} = \begin{tikzpicture}[line width=1 pt, scale=0.5, baseline=0]
\draw[line width=2 pt] (0,0)--(10,0);
\draw[densely dashed] (1,0) arc(180:0:1);
\draw (4,0) arc(180:0:1);
\draw (7,0) arc(180:0:1);
\end{tikzpicture} = \frac{(L-\tau )^{2 \eps } \tau ^{-2+\eps } \Gamma (-1+\eps )^2}{512 
\pi ^6 \Gamma (1+2 \eps )}\, \xi^{3},
\ee
\be
N_{2} = \begin{tikzpicture}[line width=1 pt, scale=0.5, baseline=0]
\draw[line width=2 pt] (0,0)--(10,0);
\draw (1,0) arc(180:0:1);
\draw[densely dashed] (4,0) arc(180:0:1);
\draw (7,0) arc(180:0:1);
\end{tikzpicture} = \frac{(L (L-\tau ) \tau )^{\eps }}{512 \pi ^6 (-1+\eps )^2 \eps ^2 \tau ^2}\,\xi^{3},
\ee
\be
N_{3} = \begin{tikzpicture}[line width=1 pt, scale=0.5, baseline=0]
\draw[line width=2 pt] (0,0)--(10,0);
\draw (1,0) arc(180:0:1);
\draw(4,0) arc(180:0:1);
\draw[densely dashed]  (7,0) arc(180:0:1);
\end{tikzpicture} = \frac{L^{2 \eps } \tau ^{-2+\eps } \Gamma (-1+\eps )^2}{512 \pi ^6 
\Gamma (1+2 \eps )}\,\xi^{3},
\ee
\be
N_{4} = \begin{tikzpicture}[line width=1 pt, scale=0.5, baseline=0]
\draw[line width=2 pt] (0,0)--(7,0);
\draw[densely dashed] (1,0) arc(180:0:1);
\draw(4,0) arc(180:0:1);
\draw(4.5,0) arc(180:0:0.5);
\end{tikzpicture} = \frac{(L-\tau )^{2 \eps } \tau ^{-2+\eps }}{1024 \pi ^6 (-1+\eps ) 
\eps ^2 (-1+2 \eps )}\,\xi^{3},
\ee
\be
N_{5} = \begin{tikzpicture}[line width=1 pt, scale=0.5, baseline=0]
\draw[line width=2 pt] (0,0)--(7,0);
\draw (1,0) arc(180:0:1);
\draw[densely dashed](4,0) arc(180:0:1);
\draw(4.5,0) arc(180:0:0.5);
\end{tikzpicture} = \frac{L^{\eps } \tau ^{-2+2 \eps }}{512 \pi ^6 (-1+\eps )^2 \eps ^2}\,\xi^{3},
\ee
\ba
N_{6} &= \begin{tikzpicture}[line width=1 pt, scale=0.5, baseline=0]
\draw[line width=2 pt] (0,0)--(7,0);
\draw (1,0) arc(180:0:1);
\draw(4,0) arc(180:0:1);
\draw[densely dashed](4.5,0) arc(180:0:0.5);
\end{tikzpicture} = 
\frac{\log (\frac{2 \tau }{L+\tau })}{512 \pi ^6 \tau ^2 \eps 
}\,\xi^{3}\lp
+\frac{1}{6144 \pi ^6 \tau ^2}\Big[ \pi ^2+12 (\log ^2 2+\log 4)+6 \log 16 \log L+6 \log 
\tau  (4+\log 4+3 \log \tau )  \lp\qquad\quad \qquad
-6 \log (L+\tau ) (4+3 \log (L+\tau 
))-12 \text{Li}_2(\tfrac{\tau }{L+\tau })\Big]\,\xi^{3}+O(\eps),
\ea
\be
N_{7} = \begin{tikzpicture}[line width=1 pt, scale=0.5, baseline=0]
\draw[line width=2 pt] (0,0)--(7,0);
\draw (1.5,0) arc(180:0:0.5);
\draw(1,0) arc(180:0:1);
\draw[densely dashed](4,0) arc(180:0:1);
\end{tikzpicture} = \frac{L^{2 \eps } \tau ^{-2+\eps }}{1024 \pi ^6 (-1+\eps ) \eps ^2 
(-1+2 \eps )}\,\xi^{3},
\ee
\be
N_{8} = \begin{tikzpicture}[line width=1 pt, scale=0.5, baseline=0]
\draw[line width=2 pt] (0,0)--(7,0);
\draw (1.5,0) arc(180:0:0.5);
\draw[densely dashed](1,0) arc(180:0:1);
\draw(4,0) arc(180:0:1);
\end{tikzpicture} = \frac{(L-\tau )^{\eps } \tau ^{-2+2 \eps }}{512 \pi ^6 (-1+\eps )^2 
\eps ^2}\,\xi^{3},
\ee
\ba
N_{9} &= \begin{tikzpicture}[line width=1 pt, scale=0.5, baseline=0]
\draw[line width=2 pt] (0,0)--(7,0);
\draw[densely dashed] (1.5,0) arc(180:0:0.5);
\draw(1,0) arc(180:0:1);
\draw(4,0) arc(180:0:1);
\end{tikzpicture} = -\frac{\log (\frac{L+\tau }{2 \tau })}{512 (\pi ^6 \tau ^2) \eps 
}\,\xi^{3}\lp
+\frac{1}{1024 \pi ^6 \tau ^2}\Big[   2 \log ^2 2+\log 16+\log 16 \log L+\log 4 \log (L-\tau )+4 \log \tau +\log 4 \log \tau \lp
+2 \log L \log \tau +3 
\log ^2\tau -4 \log (L+\tau )-\log 4 \log (L+\tau )-2 \log L 
\log (L+\tau )\lp -2 \log \tau  \log (L+\tau )
-\log ^2(L+\tau )+2 \text{Li}_2(-\tfrac{\tau }{L-\tau })-2 \text{Li}_2(-\tfrac{L+\tau }{L-\tau })\Big]\,\xi^{3}+O(\eps),
\ea
\be
N_{10} = \begin{tikzpicture}[line width=1 pt, scale=0.5, baseline=0]
\draw[line width=2 pt] (0,0)--(6,0);
\draw[densely dashed] (1,0) arc(180:0:2);
\draw(1.5,0) arc(180:0:1.5);
\draw(2,0) arc(180:0:1);
\end{tikzpicture} = \frac{\tau ^{-2+3 \eps }}{1024 \pi ^6 (-1+\eps ) \eps ^2 (-1+2 \eps )}\,\xi^{3},
\ee
\be
N_{11} = \begin{tikzpicture}[line width=1 pt, scale=0.5, baseline=0]
\draw[line width=2 pt] (0,0)--(6,0);
\draw (1,0) arc(180:0:2);
\draw[densely dashed](1.5,0) arc(180:0:1.5);
\draw(2,0) arc(180:0:1);
\end{tikzpicture} = \frac{\tau ^{-2+2 \eps } ((-1+2^{\eps }) L^{\eps }+\tau ^{\eps }-(L+\tau )^{\eps })}{512 \pi ^6 (-1+\eps )^2 \eps ^2}\,\xi^{3},
\ee
\ba
N_{12} &= \begin{tikzpicture}[line width=1 pt, scale=0.5, baseline=0]
\draw[line width=2 pt] (0,0)--(6,0);
\draw (1,0) arc(180:0:2);
\draw(1.5,0) arc(180:0:1.5);
\draw[densely dashed](2,0) arc(180:0:1);
\end{tikzpicture} = \frac{1}{512 \pi ^6 \tau ^2}\Big[ -\frac{\pi ^2}{12}+i \pi  \log 2+\frac{\log ^2 2}{2}-\log 2 \log (L-\tau )
+\log 2 \log \tau\lp
+\log \frac{2 \tau }{L+\tau }-\text{Li}_2(\tfrac{L}{L-\tau })+\text{Li}_2(\tfrac{2 L}{L-\tau })-\text{Li}_2(-\tfrac{L}{\tau })\Big]\,\xi^{3}+\mc O(\eps), 
\ea
\be
N_{13} = \begin{tikzpicture}[line width=1 pt, scale=0.5, baseline=0]
\draw[line width=2 pt] (0,0)--(6,0);
\draw[densely dashed] (1,0) arc(180:0:2);
\draw(1.25,0) arc(180:0:0.75);
\draw(3.25,0) arc(180:0:0.75);
\end{tikzpicture} = \frac{\tau ^{-2+3 \eps } \Gamma (-1+\eps )^2}{512 \pi ^6 \Gamma (1+2 
\eps )}\,\xi^{3},
\ee
\ba
N_{14} &= \begin{tikzpicture}[line width=1 pt, scale=0.5, baseline=0]
\draw[line width=2 pt] (0,0)--(6,0);
\draw (1,0) arc(180:0:2);
\draw[densely dashed](1.25,0) arc(180:0:0.75);
\draw(3.25,0) arc(180:0:0.75);
\end{tikzpicture} = -\frac{\log (\frac{L+\tau }{2 \tau })}{512 (\pi ^6 \tau ^2) \eps 
}\,\xi^{3}\lp
+\frac{1}{1024 \pi ^6 \tau ^2}\Big[ \log ^2 2+\log 16+\log 4 \log L+\log 4 \log (L-\tau 
)+\log 4 \log \tau +\log ^2\tau -\log ^2(L+\tau )\lp\qquad
-4 \log 
(\frac{L+\tau }{\tau })-2 \log (L-\tau ) \log (\frac{L+\tau }{\tau 
})-2 \log \tau  \log (\frac{L+\tau }{\tau })\lp \qquad
-2 \text{Li}_2(1-\tfrac{L}{\tau })+2 \text{Li}_2(\tfrac{-L+\tau }{L+\tau 
})\Big]\,\xi^{3}+O(\eps),
\ea
\ba
N_{15} &= \begin{tikzpicture}[line width=1 pt, scale=0.5, baseline=0]
\draw[line width=2 pt] (0,0)--(6,0);
\draw (1,0) arc(180:0:2);
\draw(1.25,0) arc(180:0:0.75);
\draw[densely dashed](3.25,0) arc(180:0:0.75);
\end{tikzpicture} = \frac{\log (\frac{2 \tau }{L+\tau })}{512 \pi ^6 \tau ^2 \eps 
}\,\xi^{3}+\frac{1}{512 \pi ^6 \tau ^2}\Big[ -\frac{\pi ^2}{12}+\frac{\log ^2 2}{2}+\log  4 +\frac{3 \log 
^2 \tau }{2}\lp
+\log L \log (\frac{4 \tau }{L+\tau })+\log \tau  
(2+\log 2-\log (L+\tau ))\lp\qquad -2 \log (L+\tau )-\frac{1}{2} \log ^2(L+\tau )-\text{Li}_2(-\tfrac{L}{\tau })\Big]\,\xi^{3}+O(\eps).
\ea
\be
D_{1} = \begin{tikzpicture}[line width=1 pt, scale=0.5, baseline=0]
\draw[line width=2 pt] (0,0)--(7,0);
\draw (1,0) arc(180:0:1);
\draw (4,0) arc(180:0:1);
\end{tikzpicture} = \frac{L^{2 \eps } \Gamma (\eps )}{64 \pi ^{7/2} (-1+\eps )^2 \eps  
\Gamma (\frac{1}{2}+\eps )}\,\xi^{2},
\ee
\be
\la{u3}
D_{2} = \begin{tikzpicture}[line width=1 pt, scale=0.5, baseline=0]
\draw[line width=2 pt] (0,0)--(5,0);
\draw (1,0) arc(180:0:1.5);
\draw (1.5,0) arc(180:0:1);
\end{tikzpicture} = \frac{2^{-7+2 \eps } L^{2 \eps }}{\pi ^4 (-1+\eps ) \eps ^2 (-1+2 
\eps )}\,\xi^{2}.
\ee

\section{{Computation of   order $\z$ term   in the one-loop beta-function \la{PS}}}

Here we shall  present some details of the  computation of the one-loop 
term in $\b_\z$  in \rf{1.3}  that were   not spelled out in \cite{Polchinski:2011im}. 
Like the $\z^3$ term  in \rf{1.3}  that follows from the scalar ladder graphs 
(cf. \rf{215}) the $-\z$  term  comes from similar    graphs  with gluon propagator
instead of the scalar one. For example, in the case of  the Wilson line   along Euclidean time $t=\tau$ 
direction    the exponent in \rf{1.1} is given by $\int d \tau ( i A_t + \phi)$   and thus 
the  gluon contribution is  minus that of the   scalar   due to   extra $i$  factor. 
This already reproduces the  expression \rf{1.3}  for the one-loop 
term in $\b_\z$. 

However,  there are  two  extra types of diagrams  that are potentially contributing at order $\z$. Their total contribution should then  be zero.  
Using the vertex renormalization method  discussed in section 2.1
 they are represented by\foot{There are, of course,  two copies of the first diagram depending on ordering of the  end-points of the  propagators.} 
\be
\begin{tikzpicture}[line width=1 pt, scale=0.5, baseline=0]
\node[above] at (2,3) {$\phi(x_{0})$}; 
\node[left] at (0,0) {$0$};
\node[right] at (4,0) {$\t$};
\draw (2,3)--(2,0);
\draw[line width=2 pt] (0,0)--(4,0);
\draw[fill=lightgray] (2,0) circle(0.75);
\end{tikzpicture}\quad = \quad
\begin{tikzpicture}[line width=1 pt, scale=0.5, baseline=0]
\node[above] at (2,3) {$\phi(x_{0})$}; 
\node[left] at (0,0) {$0$};
\node[right] at (4,0) {$\t$};
\draw (2,3)--(2,2);
\draw[line width=2 pt] (0,0)--(4,0);
\draw (2,2)--(1,0);
\draw[decorate] (2,2)--(3,0);
\end{tikzpicture}\quad + \quad
\begin{tikzpicture}[line width=1 pt, scale=0.5, baseline=0]
\node[above] at (2,3) {$\phi(x_{0})$}; 
\node[left] at (0,0) {$0$};
\node[right] at (4,0) {$\t$};
\draw (2,3)--(2,0);
\draw[fill=lightgray] (2,1.5) circle(0.75);
\draw[line width=2 pt] (0,0)--(4,0);
\end{tikzpicture}
\la{d1}
\ee
Here  the  wavy line  stands for the gauge field propagator  and the   blob  in the second diagram is  the scalar one-loop self-energy correction (given by 
the sum of the scalar-gluon loop and fermion loop).  
Below  we shall  demonstrate 
the mutual cancellation of these  two contributions  as claimed in  \cite{Polchinski:2011im}
which is  similar to  the cancellation of the non-ladder diagram contributions to the  expectation value of the circular WML   observed in \ci{Erickson:2000af}. 

Explicitly, the contribution of the first  diagram in \rf{d1}    is 
\iffa
\be
V_1 =  
\begin{tikzpicture}[line width=1 pt, scale=0.5, baseline=0]
\node[above] at (2,3) {$\phi(x_{0})$}; 
\node[left] at (0,0) {$0$};
\node[right] at (4,0) {$t$};
\draw (2,3)--(2,2);
\draw[line width=2 pt] (0,0)--(4,0);
\draw (2,2)--(1,0);
\draw[decorate] (2,2)--(3,0);
\end{tikzpicture}\quad 
\begin{array}{l}\text{\small (this is one possible order on the path} \\ \text{\small with $\phi$ coming first than $A$)} \\ 
\end{array}
\ee
\fi 
\footnote{Here $2iA(\tau_{2})\phi(\tau_{3})$ comes from the mixed term  in the expansion of $(\phi+iA)(\phi+iA)$.} 
\be
\la{H.3}
V = \frac{i}{2!}\int d^{2}\tau\ \vev{\tr \Big\{\phi(x_{0})\,\text{P}[2A(\tau_{2})\phi(\tau_{3})]\Big\}\Big(
-\int d^{4}y\ f^{abc}\ \partial_{\mu}\phi_i^{a}(y)\,A^{b}_{\mu}(y)\,\phi^{c}_{i}(y)\Big)},
\ee
where $A(\tau) = A^{a}_{\mu}\dot x^{\mu}T^{a}$ and $\phi(\tau) = \phi^{a}(x)|\dot x|T^{a}$
assuming we keep the contour  general. 
If $\tau_{2}>\tau_{3}$ this gives
\ba
& -i\,f^{a'b'c'}\tr(T^{a}T^{b}T^{c}) \int d^{2}\tau\ \vev{\phi^{a}(x_{0})A^{b}(\tau_{2})\phi^{c}(\tau_{3})\ 
\int d^{4}y\  \partial_{\mu}\phi_i^{a'}(y)\,A^{b'}_{\mu}(y)\,\phi^{c'}_{i}(y)} \lp
= -i\,f^{abc}\tr(T^{a}T^{b}T^{c})\int d^{4}y\  \int d^{2}\tau\ (\dot x^{(2)}\cdot\partial_{y}\, D(y-x_{0}))\, D(y-\tau_{3})
D(y-\tau_{2})\lp
-i\, f^{cba}\tr(T^{a}T^{b}T^{c}) \int d^{4}y\ \int d^{2}\tau\ (\dot x^{(2)}\cdot\partial_{y}\, D(y-\tau_{3}))\,  D(y-x_{0})
D(y-\tau_{2})
\ea
 and for $\tau_{2}<\tau_{3}$ we get 
 \ba
& -i\,f^{a'b'c'}\tr(T^{a}T^{b}T^{c})\int d^{4}y\  \int d^{2}\tau\ \vev{\phi^{a}(x_{0})A^{c}(\tau_{2})\phi^{b}(\tau_{3})\ 
\int d^{4}y\  \partial_{\mu}\phi_i^{a'}(y)\,A^{b'}_{\mu}(y)\,\phi^{c'}_{i}(y)} \lp
= -i\,f^{acb}\tr(T^{a}T^{b}T^{c})\int d^{4}y\  \int d^{2}\tau\ (\dot x^{(2)}\cdot\partial_{y}\, D(y-x_{0}))\, D(y-\tau_{3})
D(y-\tau_{2})\lp
-i\,f^{bca}\tr(T^{a}T^{b}T^{c}) \int d^{4}y\ \int d^{2}\tau\ (\dot x^{(2)}\cdot\partial_{y}\, D(y-\tau_{3}))\,  D(y-x_{0})
D(y-\tau_{2})
\ea
Using that $
f^{abc}\tr(T^{a}T^{b}T^{c}) = \frac{i}{4}N^{3}+... $
where dots stand for subleading terms at large $N$ 
we get 
\ba\la{dd5}
V  = & N^{3}\int d^{4}y\, \mc V(y),\\
\mc V =& -\frac{1}{4}\int d^{2}\tau\ \epsilon(\tau_{2}, \tau_{3})\, (\dot x^{(2)}\cdot\partial_{y}\, D(y-\tau_{3}))\,  D(y-x_{0})
D(y-\tau_{2})\ , 
\ea
where $\epsilon(\tau_{2}, \tau_{3})$ is the antisymmetric path ordering symbol. 
Specifying to the case of the contour being straight line  we get\foot{To simplify the analysis we  may assume  that the 
point $x_0$  also lies on the line   but far away from other points and thus not  participating in limits   leading to   short-distance singularities.} 
\be
\mc V = \frac{1}{4}\int d^{2}\tau\ \epsilon(\tau_{2}, \tau_{3})\, \frac{\partial}{\partial\tau_{3}} D(y-\tau_{3})\,  D(y-x_{0})
D(y-\tau_{2}).
\ee
Integrating by parts (using that $\frac{\partial}{\partial\tau_{3}}\epsilon(\tau_{2}, \tau_{3}) = -2\delta(\tau_{2}-\tau_{3})$)  gives 
\ba
\mc V&= \frac{1}{2}\int d^{2}\tau\ \delta(\tau_{2}-\tau_{3})D(y-\tau_{3})\,  D(y-x_{0})
D(y-\tau_{2}) \lp
 +\frac{1}{4}\int d\tau_{2}\ \left[\epsilon(\tau_{2}, \tau_{3})\, D(y-\tau_{3})\,  D(y-x_{0})
D(y-\tau_{2})\right]_{\tau_{3}=0}^{\tau_{3}=\t} \lp
= \frac{1}{2}\int d\tau'\ D(y-\tau')^{2}\,  D(y-x_{0}) 
-\frac{1}{4}\int d\tau'\  D(y-\t)\,  D(y-x_{0})
D(y-\tau')   	\lp
+\frac{1}{4}\int d\tau'\ D(y)D(y-x_{0}) D(y-\tau').
\ea
This   may be written  as 
\be
\la{H.15}
\mc V = \frac{1}{4}\int d\tau'\, [2\,H(x_{0},\tau',\tau')-H(x_{0}, \tau', \tau)+H(x_{0}, \tau', 0)].
\ee
Here  the   function $H$ is in general defined as (cf. \ci{Erickson:2000af})
\ba
\la{E.1}
H &(x^{(1)}, x^{(2)}, x^{(3)}) = \int d^{2\omega}y \,\Delta(x^{(1)}-y)\,\Delta(x^{(2)}-y)\,\Delta(x^{(3)}-y) \lp
=\frac{\Gamma(\omega-1)^{3}}{4^{3}\pi^{3\omega}} \int d^{2\omega}y\,
\frac{1}{[(x^{(1)}-y)^{2}]^{\omega-1}}\frac{1}{[(x^{(2)}-y)^{2}]^{\omega-1}}\frac{1}{[(x^{(3)}-y)^{2}]^{\omega-1}} \\
&= \frac{\Gamma(\omega-1)^{3}}{4^{3}\pi^{3\omega}}\frac{\Gamma(3\omega-3)}{\Gamma(\omega-1)^{3}}\int d^{2\omega}y\,\int ^1_0 d\alpha d\beta d\gamma\ 
\frac{\delta(1-\alpha-\beta-\gamma)\,(\alpha\beta\gamma)^{\omega-2}}{[\alpha(x^{(1)}-y)^{2}+\beta(x^{(2)}-y)^{2}+\gamma(x^{(3)}-y)^{2}]^{3(\omega-1)}}\no 
\ea
where $\Delta$ is the scalar propagator in $d=2 \omega = 4- \eps$   dimensions
(with canonical normalization).  It can   be put into the form
\ba
\la{E.5}
& H(x^{(1)}, x^{(2)}, x^{(3)}) 
= \frac{\Gamma(2\omega-3)}{4^{3}\pi^{2\omega}}\int d\alpha d\beta d\gamma\ 
\frac{\delta(1-\alpha-\beta-\gamma)\,(\alpha\beta\gamma)^{\omega-2}}{[M^{2}]^{2\omega-3}} \ , \\
& M^{2}(x^{(1)}, x^{(2)}, x^{(3)}) = \alpha(1-\alpha)\,(x^{(1)})^{2}+\beta(1-\beta)\,(x^{(2)})^{2}+\gamma(1-\gamma)\,(x^{(3)})^{2}\lp
\qquad \qquad \qquad\qquad \ \ \  -2\alpha\beta\, x^{(1)}\cdot x^{(2)}-2\alpha\gamma\, x^{(1)}\cdot x^{(3)}-2\beta\gamma\, x^{(2)}\cdot x^{(3)} .
\ea
The UV divergent contribution to \rf{H.15}   comes   only from the   limits  when two points on the line approach  each other, i.e. only  from the first $H$-function term  in \rf{H.15}.
Focussing on this  term  and setting $x_0=0$    we have 
\ba
\mc V \ \to \  &  \frac{1}{2}\int d\tau'\,H(0,\tau',\tau')\no \\
&
= \frac{N^{3}\Gamma(2\omega-3)}{2^{7}\pi^{2\omega}}\int_{0}^{\t} d\tau'\  \int d\alpha d\beta d\gamma\ 
\frac{\delta(1-\alpha-\beta-\gamma)\,(\alpha\beta\gamma)^{\omega-2}}{((\alpha+\beta)\gamma)^{2\omega-3}}\tau'^{-2(2\omega-3)}.\la{d14}
\ea
Using (\ref{F.7})  then gives for the corresponding  integral in \rf{dd5}
\be
V = N^{3}\frac{2^{-3-2 \omega } \pi ^{\frac{3}{2}-2 \omega } \t^{7-4 \omega } 
\csc (\pi  \omega ) \Gamma (-3+2 \omega )}{(-7+4 \omega ) \Gamma 
(3-\omega ) \Gamma (-\frac{1}{2}+\omega )} = \frac{N^{3}}{64\pi^{4}\t}\ \frac{1}{\omega-2}+\cdots.\la{d16}
\ee
Turning  to the scalar scalar self energy  contribution,    it can be written as  (see \ci{Erickson:2000af})
\be
S= -\frac{1}{2}\,
\frac{\Gamma^{2}(\omega-1)}{32\pi^{2\omega}(2-\omega)(2\omega-3)}
\int_{0}^{\t}d\tau'\ \frac{1}{(x_{0}-\tau')^{2(2\omega-3)}}.
\ee
Setting as in \rf{d14} $x_0=0$   we get 
\be
S = -\frac{1}{2}\,
\frac{\Gamma^{2}(\omega-1)}{32\pi^{2\omega}(2-\omega)(2\omega-3)}
\int_{0}^{\t}d\tau'\ \frac{1}{\tau'^{2(2\omega-3)}} = -\frac{\pi ^{-2 \omega } \t^{7-4 \omega } \Gamma (-1+\omega )^2}{64 
(7-4 \omega ) (2-\omega ) (-3+2 \omega )}.
\ee
Then one can check that  the contributions of the  triangular graph and  self-energy correction indeed   cancel (even exactly in $\omega$), i.e. 
\be
V + S=0.
\ee

\bibliography{BT-Biblio}
\bibliographystyle{JHEP}
\end{document}